\documentclass[12pt]{report}
\usepackage{graphicx}
\usepackage{amssymb}
\usepackage{epstopdf}
\usepackage{xspace}
\usepackage{xcolor}
\usepackage{multicol}
\usepackage{floatrow}
\usepackage{float}
\usepackage{dcolumn}
\usepackage{bm}
\usepackage[dvips]{epsfig} 
\usepackage[font=footnotesize,labelfont=bf]{caption}

\usepackage{xspace}
\newcommand{\degree}{\ensuremath{^\circ}\xspace}

\newcommand{\htp}{\ensuremath{\mathrm{H}_2^+}\xspace}

\usepackage{etoolbox}
\makeatletter
\patchcmd{\@makechapterhead}{50\p@}{-10pt}{}{}
\patchcmd{\@makeschapterhead}{50\p@}{-10pt}{}{}
\pretocmd{\subsection}{\addtocontents{toc}{\protect\addvspace{-2\p@}}}{}{}
\pretocmd{\subsubsection}{\addtocontents{toc}{\protect\addvspace{-2\p@}}}{}{}
\makeatother

\begin{document}

~~~~

\thispagestyle{empty}

~~

\vspace{-0.5in}

\begin{center}
{\Large \bf IsoDAR@KamLAND:\\~\\  A Conceptual Design Report for the\\
  Conventional Facilities \\~~\\ }
\end{center}

\begin{center}
\noindent {\it Editors: } J.R.~Alonso\footnote{Corresponding Author:
  Jose R. Alonso (JRAlonso@LBL.gov)} and K.~Nakamura \\for the
IsoDAR Collaboration
\end{center}

\vspace{0.25in}

{ \it Abstract:}  

This Conceptual Design Report addresses the site-specific
issues associated with deployment of the IsoDAR experiment
at the KamLAND site.  This is the second volume in the IsoDAR Conceptual Design Reports, where the first volume was ``IsoDAR$@$KamLAND: A Conceptual Design Report for the Technical Facility,''
arXiv:1511.05130.

This document describes requirements
for the caverns to house the cyclotron, beam transport line,
and target systems; issues associated with transport and assembly
of components on the site; electrical power, cooling and ventilation; 
as well as issues associated with radiation protection of the
environment and staff of KamLAND who will be interfacing
with IsoDAR during its operational phases.
Specifics of IsoDAR operations at the KamLAND site 
are not addressed.

Recent developments in planning for deployment of IsoDAR include the
identification of a potential new site for the experiment, where the target can be placed
directly on the equatorial plane of the KamLAND detector, and also,
an upgrade of the 
detector resolution to 3\%/$\sqrt{E(MeV)}$.   The option of the new site
might allow, depending on the results of shielding and background evaluations in KamLAND, for an increase
in event rate by about a factor of 1.6 owing to increased
solid angle for the detector, improving the physics reach for the same period of the experiment.
Alternatively, it raises the option of reducing technical risk and cost by reducing beam intensity to
maintain the originally planned event rates.    This
new siting option is described, and aspects the physics reach of the sterile neutrino search
are updated to reflect this second option, as well as the higher resolution of the experiment.
A full update of the physics capability given the new site and resolution is beyond the
scope of this CDR and will be published later.



\newpage

~~~~~

\vspace{1.in}

\begin{center}
 {\it \Large Contributing Authors: }\\
~~~\\
~~~\\

A. Adelmann$^6$
J. Alonso$^5$,
L.~Bartoszek$^{1}$, 
A.~Bungau$^{3}$,
L.~Calabretta$^{4}$,  
J.M.~Conrad$^{5}$,
S.~Kayser$^{5}$, 
M.~Koga$^{8}$,
K.~Nakamura$^{8}$,
H.~Okuno$^{7}$,  
M.H.~Shaevitz$^2$, 
I.~Shimizu$^{7}$, 
J.~Smolsky$^{5}$,
K.~Tanaka$^{7}$,
Y.~Uwamino$^{7}$,
D.~Winklehner$^{5}$, 
L.A~ Winslow$^{5}$.  

~~~~~\\

{\it 
$^1$Bartoszek Engineering, Aurora IL, US\\
$^2$Columbia University, New York NY, US\\
$^{3}$University of Huddersfield, Huddersfield, UK\\
$^{4}$INFN  Sezione di Catania, IT \\ 
$^{5}$Massachusetts Institute of Technology, Cambridge MA, US\\
$^{6}$Paul Scherrer Institute, Villigen, CH\\
$^{7}$RIKEN Nishina Center for Accelerator-based Science, Wako,  JP\\
$^{8}$Tohoku University, Sendai, JP\\}
\end{center}

\newpage 

\tableofcontents

\clearpage
\chapter{Introduction:  IsoDAR Experiment Overview}

IsoDAR (Isotope Decay At Rest) is a
novel, pure $\bar \nu_e$ source that, when paired with KamLAND allows for state-of-the-art
Beyond Standard Model searches for sterile neutrinos \cite{PRL} and
beyond Standard Model interactions \cite{elastic}.  

The experiment makes use of a cyclotron-accelerated beam delivered to a decay-at-rest target.
Specifically, a high-intensity H$_2^+$ ion source feeds  a 
60 MeV/amu cyclotron.   After acceleration,  the extracted H$_2^+$
ions are stripped to form a proton beam.
The proton beam is then transported to a
$^9$Be target, producing neutrons.  The neutrons enter 
a $\ge$99.99\% isotopically pure $^7$Li sleeve, where neutron
capture results in $^8$Li.  The  $^8$Li isotopes undergo
$\beta$ decay-at-rest to produce an isotropic, very pure
$\bar \nu_e$ flux.
Pairing this  very high-intensity $\bar \nu_e$ source with a detector that contains hydrogen, 
such as KamLAND, allows physics searches which 
exploit the inverse beta decay (IBD) interaction, $\bar \nu_e
+ p \rightarrow e^+ + n$ and  $\bar \nu_e$-$e^-$ elastic scattering
processs.

This document addresses the site-specific issues associated with deployment
of the IsoDAR neutrino source at the KamLAND site.   
The ``technical facility,'' which describes the design and
construction of the neutrino source, including the accelerator and the
target/sleeve, is described in a separate CDR \cite{techCDR}  (arXiv:1511.05130v1).    

\section{Physics}

A detailed description of physics goals for IsoDAR is presented
in Chapter 2 of the CDR for the Technical Facility
(arXiv:1511.05130v1).  Those physics goals are described within the
context of the layout called Site 1 (the original site) in this CDR.    Specifically,
the target/sleeve center is located 16 m from the center of the
KamLAND detector.     This document introduces Site 2 (now our preferred site), a new location
on the opposite side of KamLAND which
allows the target/sleeve center to potentially be about 12.5 m from the center of the
KamLAND detector.    

Because of its proximity, Site 2 represents a $\times 1.6$ increase in
flux.  This leads to a choice for the experiment.   The facility can
be run at full power (5 mA of H$_2^+$ providing 10 mA of protons), in which case the physics
goals can be accomplished faster.    Alternatively, the facility can be run at a
factor of $\times 1.6$ less power (3 mA of H$_2^+$ providing 6 mA of
protons),  which reduces technical risk and operations cost, for the originally
planned run-length.      In the discussion below, we assume the statistics and time-period of
the 16 m scenario (that is, we consider the Preferred Site run at $\times 1.6$
less power).

Since the original development of the IsoDAR physics case,  KamLAND has
also introduced a plan to upgrade the detector that will improve the
resolution from 6.4\%/$\sqrt{E}$ to 3\%/$\sqrt{E}$.   We briefly
consider the expected improvement from this upgrade in the discussion below.

\subsection{Sterile Neutrino Search}

IsoDAR was developed to allow a highly sensitive sterile
neutrino search.  A sterile neutrino is a lepton that has no standard model interactions
but that can mix with the regular neutrinos involved in the weak interaction.   
Searches for light sterile neutrinos with mass $\gtrsim$1~eV are motivated by observed
anomalies in several experiments.    IsoDAR provides a uniquely
sensitive method to search for sterile neutrinos through
reconstruction of the $L/E$ dependence of the neutrino oscillation process,
commonly referred to as an ``oscillation wave.''    

Fig.~\ref{IsoFig}, red, shows the sensitivity of IsoDAR$@$KamLAND for
the original configuration, with the target at 16 m.    Because the
oscillation signal depends on $L/E$, if we consider the new site, at 12.5 m,
for the same statistics, 
the sensitivity will become slightly worse at low mass splittings and
slightly better at high mass splittings.  This can be seen by the blue
line.  At very high mass splittings, the improved energy resolution
also improves the reach.

A useful review of the design of sterile neutrino experiments and the global
comparisons used to search for sterile neutrinos  
is provided in Ref.~\cite{sterilechapter}.  Current global fit results also 
appear in Refs.~\cite{SBL2012, SBL2016, IceFits} and are discussed below.

\subsubsection{The Status of Sterile Neutrino Searches Today}

\begin{figure}[t]
\begin{center}

{\includegraphics[width=4.in]{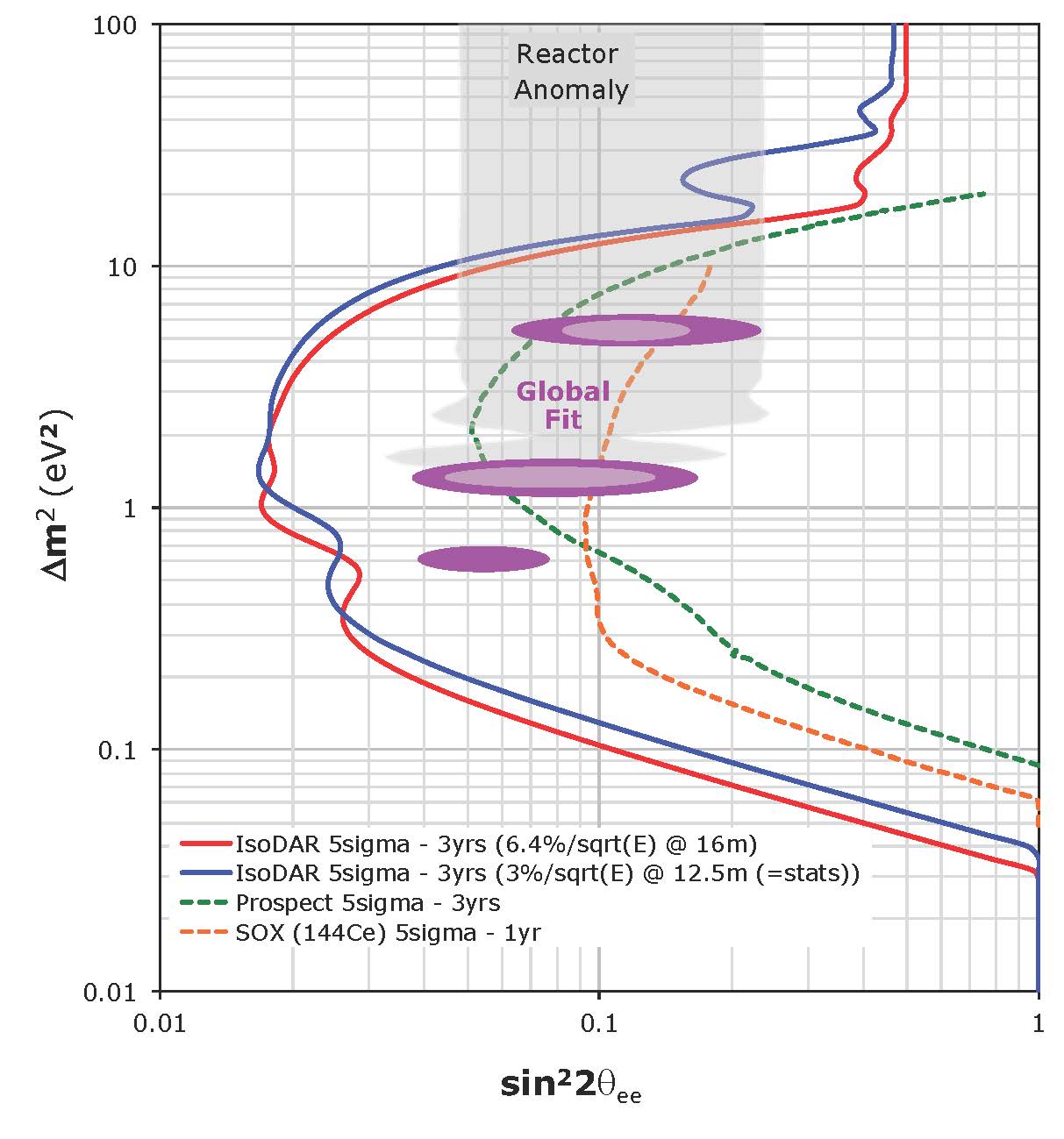}} 
\end{center}
\caption{\footnotesize  The solid curves indicate $\Delta m^2$ vs. $\sin^2 2\theta_{ee}$ 
5$\sigma$ exclusion limit reach with IsoDAR@KamLAND
(three-year run): Red--16 m (Site 1, the original location, 10 mA of proton current); Blue--12.6 m (Site 2, the now 
preferred site, 6 mA
proton current).   The
allowed regions for electron-disappearance anomalies (grey) and the
sensitivity of other upcoming experiments (green, orange dashed) are explained in the
text.   The allowed regions
  at 99\% CL from a 2016 global fit
  as a function of  mass splitting, $\Delta m^2$, vs. electron-flavor
  disappearance mixing angle, $\sin^2 2\theta_{ee}$  are shown in
  magenta \cite{IceFits}.   
\label{IsoFig}}
\end{figure}

Anomalies that point to the existence of sterile neutrinos come from a
wide range of experiments employing neutrinos and antineutrinos of
different flavors and different energies \cite{LSND, MBnu, MBnubar,
  reactor1, reactor2, SAGE3, GALLEX3}.      The
significance of these anomalies range up to 4$\sigma$, and
taken together, point to a new oscillation parameter region 
that can be
explained by introducing a new sterile neutrino type.  
This model depends on 
four additional parameters, $\sin^2 2\theta_{ee}$,
 $\sin^2 2\theta_{\mu\mu}$,  $\sin^2 2\theta_{\mu e}$, and $\Delta
 m^2$. The first three are the mixing angles measured in three
 types of oscillation experiments
 $\nu_e$ disappearance, $\nu_\mu$ disappearance and
$\nu_\mu \rightarrow \nu_e$ appearance respectively. 
With one new sterile neutrino type, the squared mass
 splitting between the mostly sterile and mostly active states must be consistent for all three types of oscillations.
 In contrast to the experiments with anomalies, other experiments
 potentially sensitive to sterile neutrinos
have observed null results  \cite{KARMEN, ConradShaevitz, MBNumi,
  MB_SB, MBnudisapp, NOMAD1, CCFR84, CDHS, IceCubePRL}, and, thus, limit the parameter space
of the anomalies.    The resulting 99\% CL allowed parameter region
for fits to the entire data set,  as a function of $\Delta m^2$ and
$\sin^2 2\theta_{ee}$ , is shown on  
Fig.~\ref{IsoFig} (magenta)   (See  Ref.~\cite{IceFits} for plots with respect to $\sin^2 2\theta_{e\mu}$ and 
$\sin^2 2\theta_{\mu\mu}$.)
The results are consistent with a sterile neutrino model with a
sterile--active mass squared mass difference of $\gtrsim$1 eV$^2$. 

The global fit allowed regions have changed very little recently, as seen by comparing Ref. \cite{IceFits} to \cite{SBL2016},
despite a great deal of press given to null results published in spring/summer of 2016
\cite{IceCubePRL, MINOS2016, An:2016luf, MINOSDAYABAY}.   The
stability arises because the global fit result is highly significant.
It is a $>6\sigma$ improvement over the
no-sterile-neutrino hypothesis ($\Delta \chi^2_{null-min}$/dof of
50.61/4) and is located at the edge, in parameter space, of the reach of
most null experiments including  those associated with Refs.~\cite{IceCubePRL, MINOS2016, An:2016luf, MINOSDAYABAY}.
It should be noted that these experiments do not actually claim to cover the
full global fit region; they only claim to cover the highest
confidence level allowed region of  $\nu_\mu \rightarrow \nu_e$
experiments, omitting the disappearance signals.   While focusing on a
chosen subset is questionable, this does, however, draw attention to an important
point:  the preferred global allowed region is located at the extreme
edge in mixing angle of
the appearance-only allowed region, pulled by the disappearance results.  This ``tension''
between appearance and disappearance results could come about due to systematic
effects in one or more data sets or from a problem with the sterile
neutrino model.   This well-known tension is one of the
primary reasons scientists are uncomfortable with the sterile neutrino
model, despite the  $>6\sigma$ improvement in the fits when the model
is introduced.

The global fits are also claimed to be in tension with cosmological data.     However, results from the CMB
and large scale structure (LSS) constraints are model dependent.    Reasonable examples of models
that evade the cosmological limit are cited in Refs.~\cite{IceFits, sterilewhitepaper}.   Based on this, 
it is most interesting to fit the cosmological data separately 
from the oscillation experiments, and then consider the meaning of discrepancies.

To address these results,  we
need a definitive
experiment with very high significance, with low systematic effects,
and with ability to reconstruct the  $L/E$ dependence
that is the signature for oscillations.

\begin{figure}[t]
\begin{center}
\includegraphics[width=.40\textwidth]{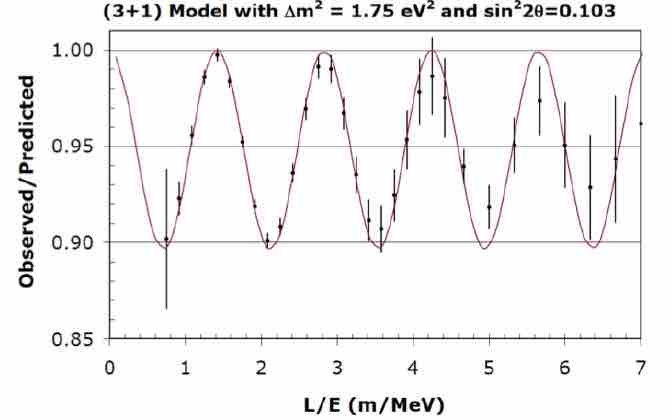}~\includegraphics[width=.39\textwidth]{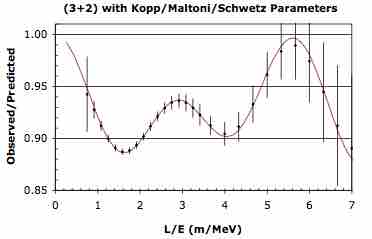}
\end{center}
\vspace{-.5cm}
\caption{\footnotesize IsoDAR$@$KamLAND $L/E$ dependence, 5 years of running,
  for one (left) \cite{SBL2016} and two
  (right) \cite{kopp}  sterile neutrinos.   Solid curve is the oscillation probability with no smearing in the reconstructed position and energy
  and the data points with error bars are from simulated events including smearing.
\label{osc3N}}
\end{figure}

\subsubsection{The Unique Power of IsoDAR for Sterile Neutrino
  Searches}

IsoDAR addresses the important design points for
mounting a definitive, next-generation search for sterile neutrinos. 
Specifically, IsoDAR has very high statistical precision with small
systematic uncertainty. 
The single, high-endpoint energy spectrum associated with the $^8$Li  decay provides a
very pure, isotropic $\bar \nu_e$ flux with known energy and angular distribution.  
Due to the target and detector shielding, the backgrounds from the neutrino source are expected to be low.  
Background effects are further reduced by exploiting the $L/E$ dependence associated with an oscillation signal. 
Non-beam backgrounds can be monitored and subtracted using beam off running.
The relatively small neutrino production region in the source and the excellent KamLAND event reconstruction capability
lead to IsoDAR's excellent precision in searching for sterile neutrino oscillations and mapping out the
``oscillation wave'' .

Fig.~\ref{osc3N} illustrates 
the power of the experiment to go beyond other proposals, showing the 
ability to differentiate models that have one sterile neutrino (left) 
and two sterile neutrinos (right) via measurements of the
``oscillation wave'' \cite{PRL}.  
This, in turn, yields the sensitivity shown on Fig.~\ref{IsoFig},
which definitively addresses the electron-flavor disappearance anomalies (grey)
\cite{reactor}  and the global fit 99\% CL allowed regions
\cite{IceFits} (magenta).

Fig.~\ref{IsoFig} compares the 5$\sigma$ capability of IsoDAR, a ``driven source experiment''
(the decaying isotope is constantly replenished by an external beam), compared to a future
non-driven source experiment (SOX) \cite{SOX}.      Neutrino sources that are not ``driven'' tend to have low
endpoint neutrino energy and also short lifetimes.    The SOX $^{144}$Ce$\rightarrow$$^{144}$Pr$\rightarrow$$^{144}$Nd
source will produce neutrinos up to $\sim 3$ MeV and will be located below the Borexino
detector at $L=8.25$ m from the center.   The half-life of this source
is 296 days, and so, as a result, they expect only a 1 year run.

Fig.~\ref{IsoFig} shows the 5$\sigma$ sensitivity for a future reactor experiment,
PROSPECT. This is one of as many as five reactor experiments that
may present new results by the end of 2018 \cite{manyreactor}. 
Two reactor experiments that have already presented first
results,
NEOS \cite{NEOS} and DANSS \cite{DANSS, manyreactor}
results have spectral features that agree with 
a sterile neutrino interpretation and enhance the global fit \cite{Schwetzreact}.
Reactor fluxes have two important issues that the IsoDAR flux avoids.  The first is an excess in the reactor spectrum at 5 MeV
that is seen in most \cite{DC5MeV, RENO5MeV, DB5MeV}, but not all \cite{Bugey}
reactor experiments, for which the source is unknown.    Second, Daya Bay has
indicated the reactor anomaly may depend on fuel (U vs. Pu) \cite{DBflux}, however
three independent analyses \cite{Schwetzreact, Hayesreact,
  Giuntireact} reach the conclusion that even after the new Daya Bay
result, the reactor anomaly
remains at $>2\sigma$. Quoting Ref.~\cite{Schwetzreact}, ``We find that the sterile neutrino hypothesis cannot be rejected based
on global data and is only mildly disfavored compared to an individual rescaling of
neutrino 
fluxes from different fission isotopes.''
IsoDAR does not face these questions. 
The well-understood, high-endpoint $^8$Li  decay spectrum provides an
isotropic $\bar \nu_e$ flux with known energy
distribution.  

IsoDAR$@$KamLAND not only covers the existing 3+1-fit parameter space, it extends
comfortably beyond that space.     As a result, if the anomalies
shift, or if new models for sterile neutrino oscillations are
developed, there is still space for this experiment to yield a
decisive result.

\subsection{NSI Search Through Precision Electroweak Tests}

\begin{figure}[t]
	\begin{center}		
		\includegraphics[width=0.39\textwidth]{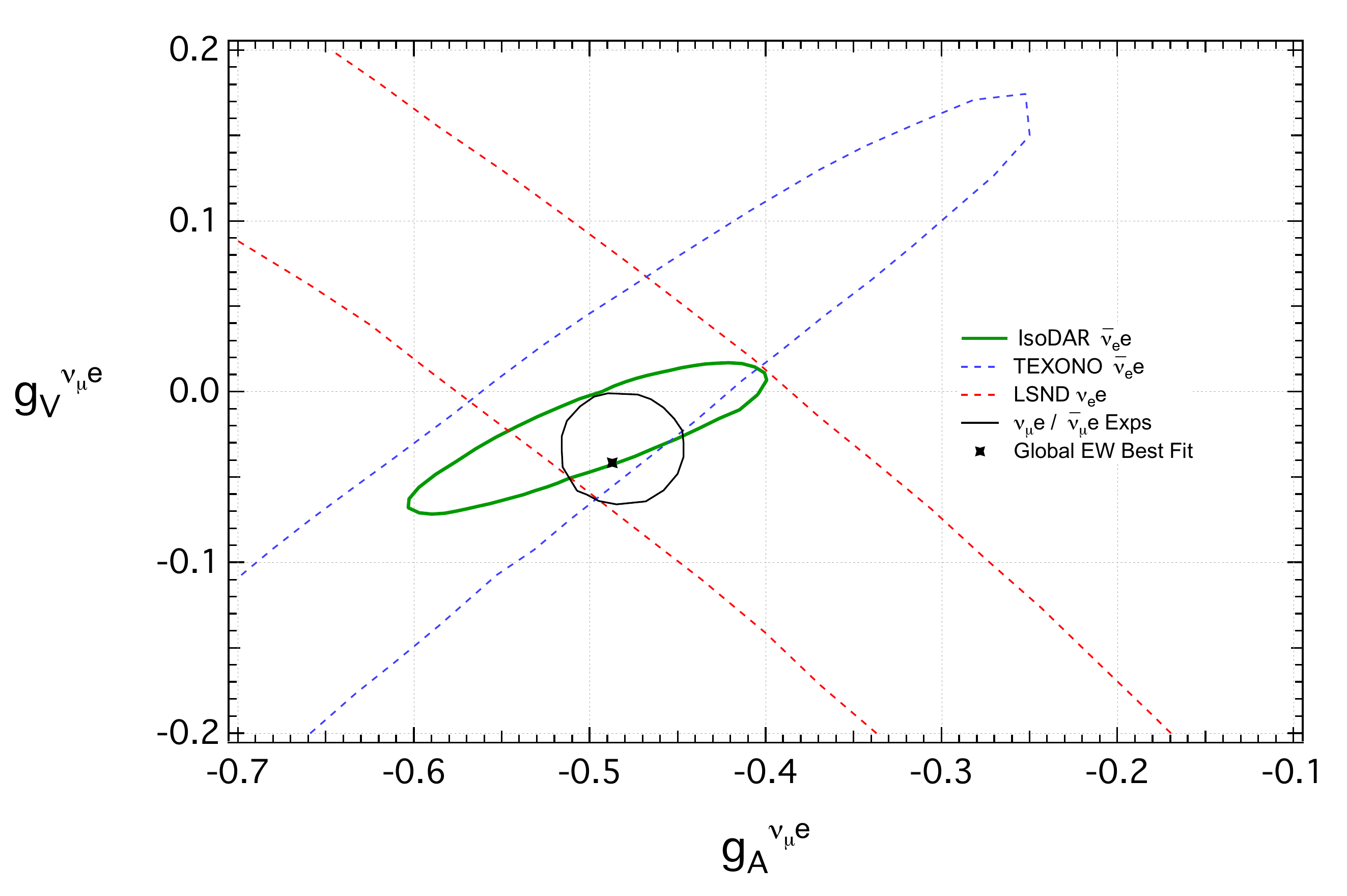}
                \vspace{-0.4in} ~ {\includegraphics[width=2.5in]{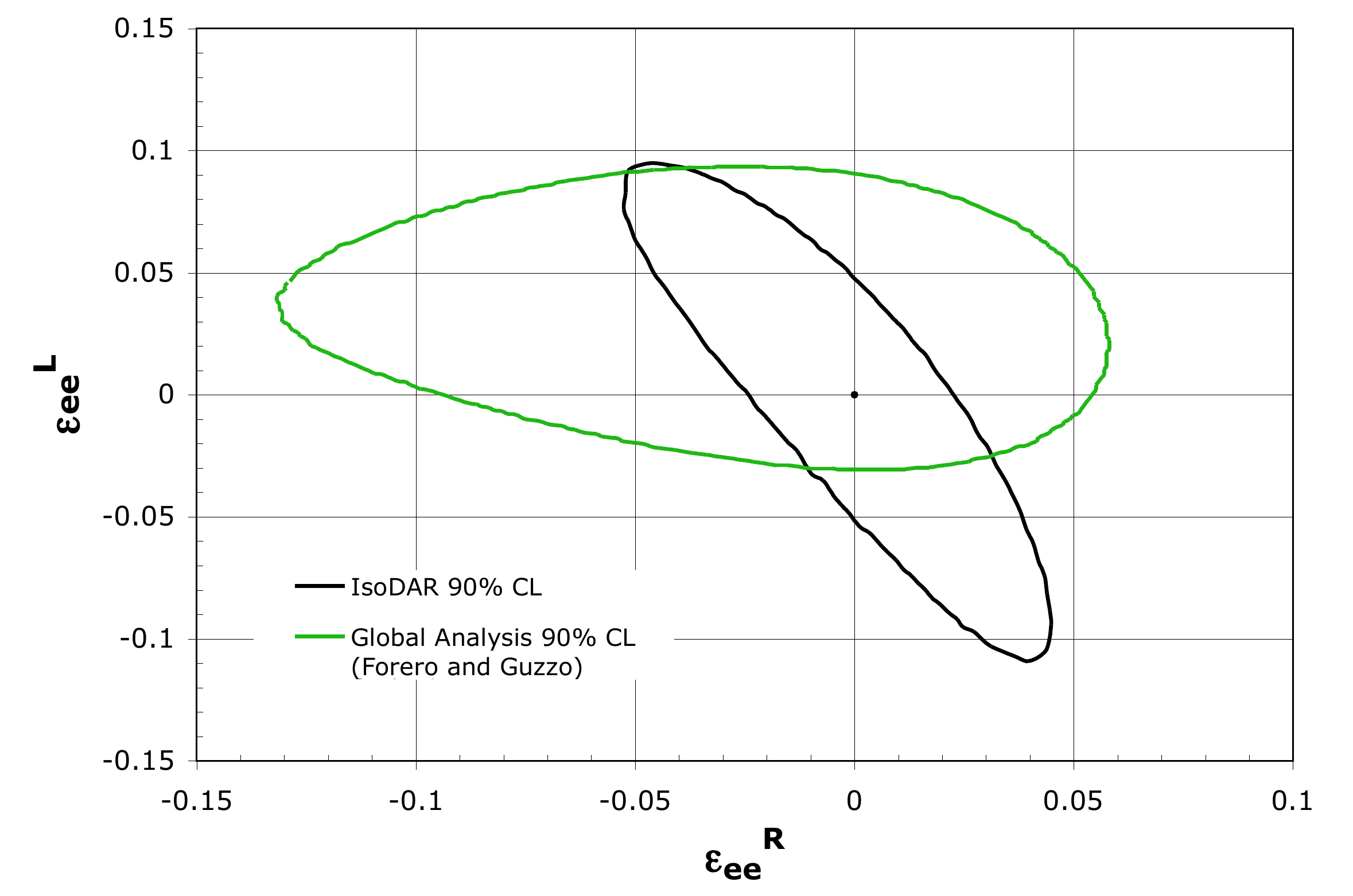}}
	\end{center}
	\vspace{0.2in}
	\caption{\footnotesize\label{gV_gA} Left:   1$\sigma$
          sensitivities and results for $g_V$ and $g_A$ and the
          electroweak global best fit point \cite{nakamura:pdg}, where $g_{V,A}^{\nu_\mu e}=
          g_{V,A}^{\nu_e e}-1$ is assumed. The $\nu_\mu
          e/\bar{\nu}_\mu e$ contour is at 90\% C.L.
          Right: Comparison of the NSI
  sensitivity to the global limit }
	\end{figure}

The search for indications of non-standard neutrino interaction effects, NSIs, is a second example of 
the outstanding new physics opened up by the IsoDAR
source.      This makes use of the fact that with a 1 kiloton
detector, such as KamLAND, the experiment will collect the largest sample of low-energy $\bar
\nu_e$-electron elastic scatters (ES) that has been observed to date
\cite{elastic}.    The ES rate can be normalized
using the well-understood, very-high-statistics IBD events.

Approximately 2600 ES events will be collected at KamLAND above a 3~MeV visible
energy threshold over a 5 year run--a factor of five above existing $\bar
\nu_e$ 
samples \cite{elastic}.      The higher resolution of the KamLAND
detector upgrade is likely to allow the 3-MeV threshold to be
lowered.   That will substantially benefit the statistics, since the
visible energy distribution for ES scattering is peaked at low energy
or low $y$,
where $y=E_{e^-}/E_{\nu}$.      This improvement will be
investigated in the future.

Ref.~\cite{elastic} describes how 
the ES cross section can be used to measure the weak couplings, $g_V$ and $g_A$, as well as $\sin^2\theta_W$, a fundamental parameter of the Standard Model as described in Ref.~\cite{nakamura:pdg}.  Although $\sin^2\theta_W$ has been determined to high precision~\cite{baak:electroweak}, there is a longstanding discrepancy~\cite{nakamura:pdg} between the value obtained by $e^+e^-$ collider experiments and the value obtained by NuTeV, a precision neutrino-quark scattering experiment
\cite{zeller:nu-nucleus_scattering}.  
Despite having lower statistics than NuTeV, IsoDAR would measure
$\sin^2\theta_W$ using the purely leptonic ES interaction, which is well understood theoretically and does
not involve any nuclear corrections from theoretical models.
To compare the sensitivity of IsoDAR with that of other experiments,
the fits to the ES cross section can be done in terms of $g_V$
and $g_A$, as shown in Fig.~\ref{gV_gA}, left.  
We find that IsoDAR is the
only experiment to date that can test the consistency of $\nu_e e/\bar\nu_e e$ couplings with $\nu_\mu e/\bar\nu_\mu e$.

The ES cross section is also sensitive to new physics in the neutrino
sector arising from nonstandard interactions (NSIs), that give rise to
corrections to $g_V$
and $g_A$.   These two couplings can be recast  in terms of the left
and right handed couplings, that may be modified by NSI's:  $g_L^{SM} \rightarrow g_L^{SM} +
 \varepsilon^{eL}_{ee}$ and/or $g_R^{SM} \rightarrow g_R^{SM} +
 \varepsilon^{eR}_{ee}$.  The IsoDAR sensitivity is stronger than, and
orthogonal to, the existing global limit \cite{elastic}, as shown in
Fig.~\ref{gV_gA}, right.

\subsection{Other Physics}

IsoDAR also makes other beyond-Standard Model searches.   Two examples of
beyond-Standard Model studies  under development are: 1) A search
for rare BSM decays of $^8$Li \cite{rareLi} through an accurate measurement of the
neutrino flux.  This can be pursued if sterile neutrino oscillations are not observed.   This is a complementary approach
to the ion trap  $^8$Li-decay experiments now underway \cite{iontrap}.
2) A search for a BSM signal in trident events, $\bar \nu_e + C
\rightarrow \bar\nu_e e^+ e^- + C$.  Preliminary estimate is that
more than 100 Standard Model trident events will be produced.    
It can also 
provide nuclear physics measurements
and it provides calibration for KamLAND. 


\section{Technical Components}

The steps in generating the antineutrinos of interest are as follows:

\begin{itemize}
\itemsep-0.1em
\item A cyclotron produces a beam up to a maximum of 5 milliamperes of H$_2^+$ (10 mA of protons)
at 60 MeV.
\item A transport line that includes a stripper section which converts the H$_2^+$ to protons.
\item A beryllium target that is struck by the proton beam, producing large quantities of
neutrons.
\item A sleeve containing a mixture of highly-enriched ($>99.99\%$) $^7$Li and beryllium that 
is flooded by these neutrons, which are moderated and multiplied by the beryllium
and are captured by the $^7$Li to make
the parent $^8$Li.
\item Shielding of all components that inhibits neutrons escaping into the environment.
\end{itemize}

\subsection{Cyclotron System}

\begin{figure}[t]
\centering
\includegraphics[width=5in]{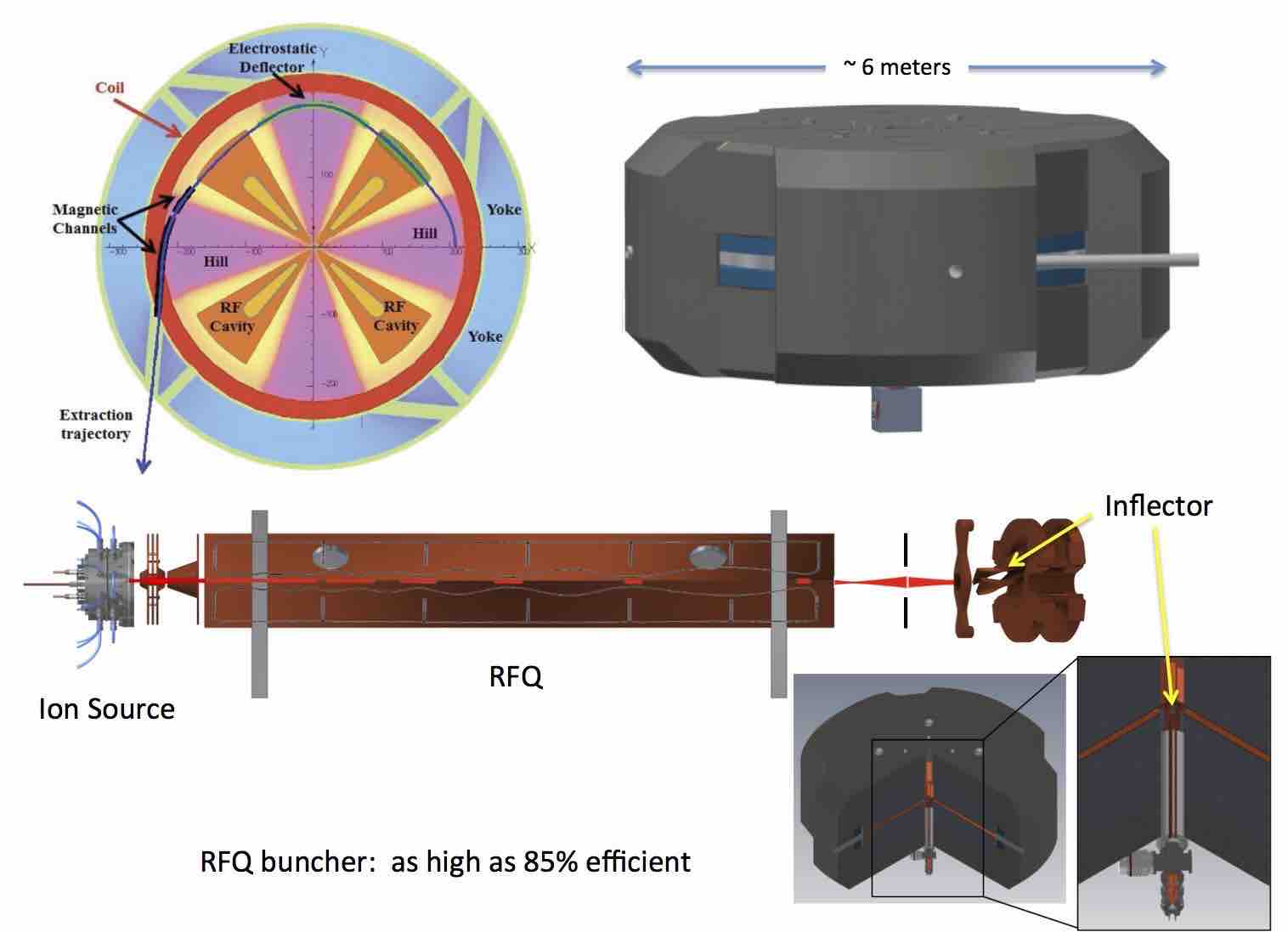}
\caption{{\footnotesize Schematic of cyclotron system elements.  Upper right is the outside of the steel,
mainly the 400 ton yoke of the magnet.  
The tube extending tangentially in the horizontal plane is the extraction port, 
the full energy beam emerges here.  The protrusion below the center is the injection line, 
shown in more detail in the insets at lower right.  
The ion source is a multi-cusp, filament driven high-current source of H$_2^+$  ions,
 that are sent through the RFQ (Radio Frequency Quadrupole) pre-accelerator 
 that provides a small energy boost, but more importantly bunches the beam 
 at the cyclotron frequency of about 50 MHz.  This bunching allows capture 
 efficiency of the ions to be almost 85\%.  This bunching scheme is a new feature of this
 accelerating system.  
 The insets show the RFQ inserted in a hole bored in the center of the cyclotron steel, its exit
 must be as close as possible to the RF accelerating electrodes to not lose the tight bunching
 from the RFQ.  A spiral inflector channels the beam from its axial orientation to bring it into the
 plane of the cyclotron.
 The top left section shows the internal magnetic structure with four ``hills'' (pink) 
 and four RF dee's.  The beam spirals out, gaining energy, until it reaches the extraction channel.}
\label{cyclotron}}
\vspace{0.2in}
\end{figure}

Figure~\ref{cyclotron}
shows the main components of the beam-forming elements:  the ``Front End'' (ion source and 
RFQ pre-injector), the spiral inflection system that brings the beam into
the central plane of the cyclotron, the internal view of the cyclotron including the
``hills'' and ``valleys'' of the magnet (where the accelerating dee's are located in the valleys),
and the extraction channel with its electrostatic and magnetic
elements.

There is a great deal of equipment associated with this system.   The
Technical Facility CDR provides details on the equipment.    In order
to bring this equipment into the KamLAND space, the size of the pieces
will be limited (see discussions in later chapters).  The Technical
Facility CDR explains how the pieces associated with this equipment
will be broken down for transport to the area.

The current required, 10 mA of protons, is higher than what is available from existing
cyclotrons.   For a discussion of the challenges involved in
accomplishing this, see the Technical Facility CDR.    Briefly, we
have introduced a number of innovations to address the problems of
running at very high current.    These include acceleration of 
H$_2^+$ ions: a hydrogen molecule with one electron removed.  
Two advantages of using this ion are:  there are two protons for every charge, 
so the Coulomb forces are reduced, and also, the heavier ion (twice as massive
as a single proton) presents more inertial mass.  These also include 
pioneering a new bunching scheme using an RFQ accelerator, 
which has the potential for achieving 85\% bunching efficiency. 

\subsection{Beam Transport (MEBT) System}

The H$_2^+$  ions at 60 MeV/amu extracted from the cyclotron are transported
to the target via the MEBT (Medium Energy Beam Transport). 
The H$_2^+$  ions are passed through a thin
carbon stripper foil that removes the electron, leaving two bare protons.
It is easier and safer to transport protons; beam losses can be better controlled.
The plan is to place the stripper and an analyzing magnet that separates protons from
H$_2^+$  inside the main cyclotron shielding.  This will be the highest beam-loss point
in the MEBT, but the shielding is sufficient to contain any stray neutrons.

Beyond the stripper stage, the beam is transported through to the target area.  
Standard transport elements are used:  quadrupole magnets for focusing and dipoles
for bending the ions.  The evacuated pipe through which the beam runs is held at
a very high vacuum to minimize losses due to scattering with gas atoms.
Though substantially less for bare protons than H$_2^+$, gas scattering is still a 
source of activation from protons escaping the beam trajectory
for the very intense beams we are producing.
The ultra-high vacuum
Very little beam loss is anticipated in this area, so plans call for minimal shielding.
However should simulations and safety considerations
 indicate that more shielding is needed, cavern
designs must accommodate this.

As the beam enters the target region it first goes through a 30\degree magnet that provides
protection from neutrons emitted in the backward direction from the target.   
A dump for these neutrons adds to the shielding requirements in the vicinity of the target, which
also must be taken into consideration in specifications for cavern dimensions.
The last part of the MEBT can be seen in Figure~\ref{tgt3d} in the
next section.  Further details of the MEBT equipment are discussed in
the Technical Facility CDR.

\subsection{Target System}

\begin{figure}[t]
\centering
\includegraphics[width=5in]{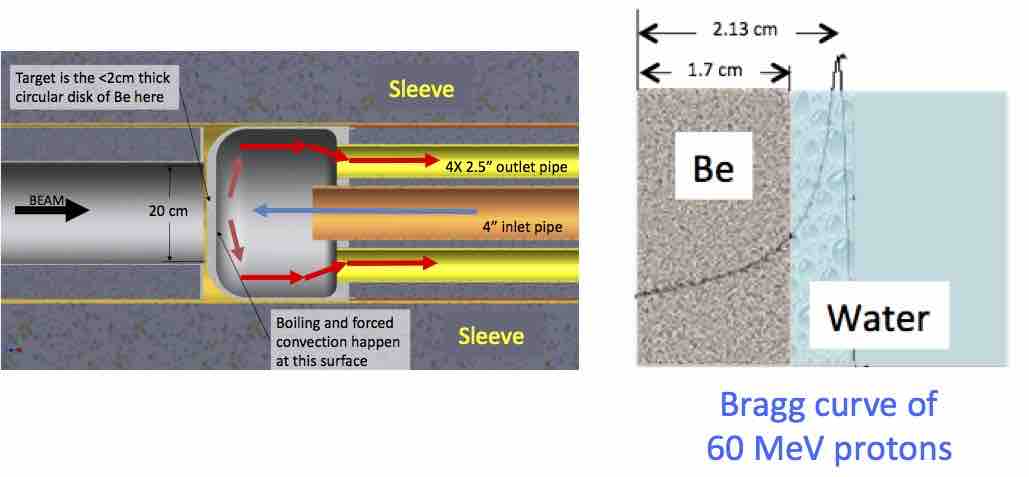}
\caption{{\footnotesize  Inside of the target vessel.  The beam is spread out over the face
of the target by upstream wobbler magnets running at least at 50 Hz.  The 60 MeV protons
produce neutrons in the target, the overall efficiency is about 10\% (1 neutron for every 10 protons).
Cooling is provided by circulating heavy water, that also serves to moderate the neutron energies. 
The protons actually stop in the water, depositing their Bragg Peak energy there.  Neutrons stream
into the sleeve surrounding the target, consisting of a mixture of highly enriched ($>$99.99\%) $^7$Li 
and beryllium.  $^8$Li is produced by neutron capture.}
\label{target-closeup}}
\vspace{0.2in}
\end{figure}

Figure~\ref{target-closeup} shows the details of the target.  The proton beam
strikes a beryllium disk, 20 cm diameter, and
1.6 cm thick.  As the inset shows, the range of the 60 MeV proton beam is 2.1 cm
in beryllium, so the beam actually stops in the heavy-water cooling bath behind the 
target.  As the neutron production efficiency falls off with lower energy (but the rate
of energy deposition increases), stopping the beam directly in the cooling water decreases
the heat load of the target without appreciably affecting the neutron yield.

Heavy water is used as the cooling fluid, as the absorption cross section for neutrons is
substantially lower, increasing the neutron flux reaching the $^7$Li in the sleeve.  Using 
heavy water does introduce tritium issues, however the technology
for controlling this is well understood, and has been incorporated into the planning for 
the primary cooling loop.  
This water flows at high pressure onto the back face of the target, 
producing turbulent flow of the
fluid, which is optimal for heat transfer from the beryllium.
Modeling studies of the heat transfer process includes this turbulence, as well
as extra heat deposited by the protons stopping in the liquid.  Boiling and 
bubble formation are important elements of this modeling process.

\begin{figure}[t]
\centering
\includegraphics[width=5in]{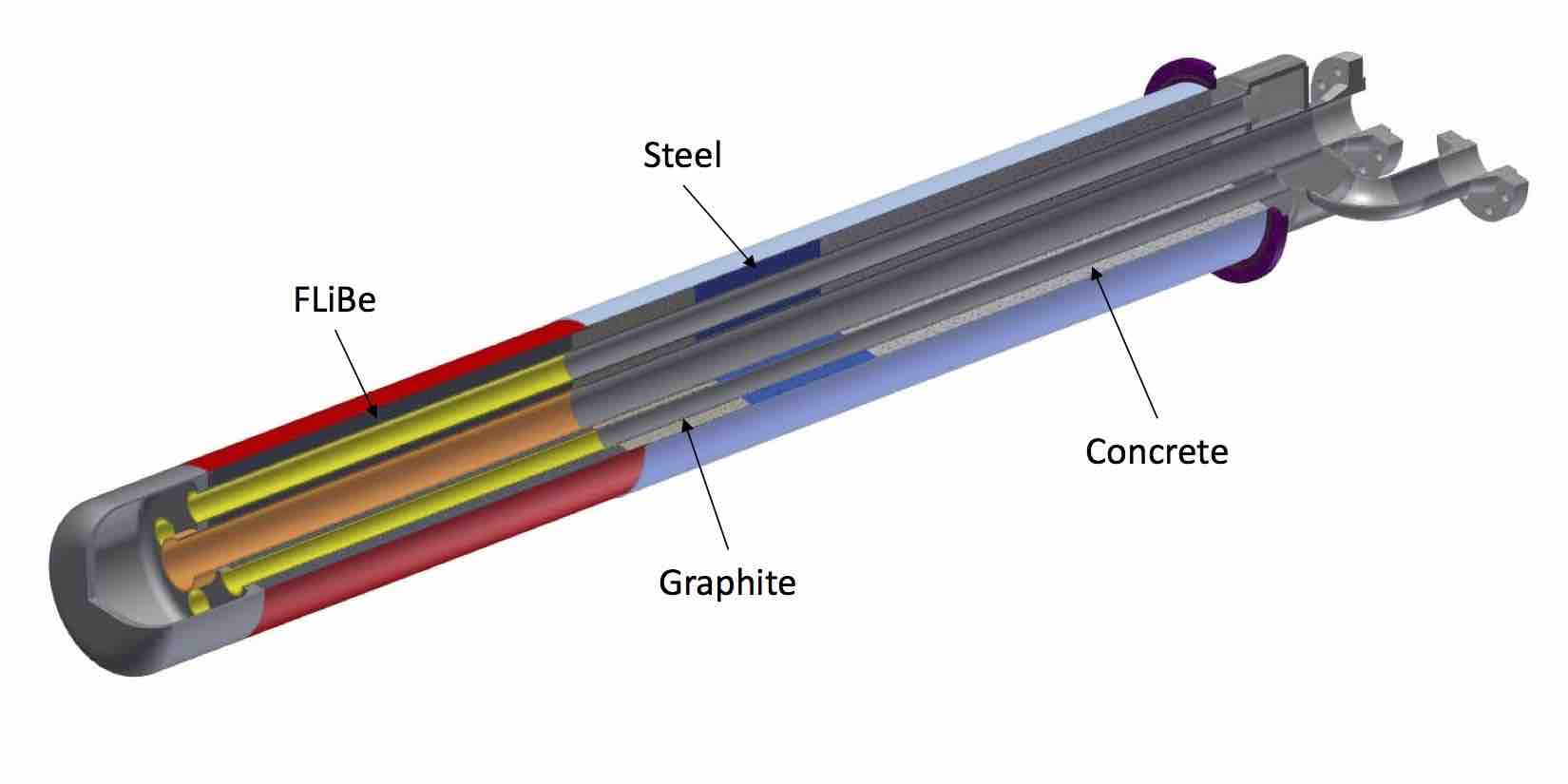}
\caption{{\footnotesize  Referred to as the ``Torpedo,'' the target assembly consists
of the beryllium surface hit by the beam, as well as the cooling system.  The entire assembly
can be easily detached from the cooling lines and removed into a shielded casket, as seen in 
Figure~\ref{TgtPlanView}}
\label{torpedo}}
\vspace{0.2in}
\end{figure}

The target assembly is referred to as the ``target torpedo.''   (See Figure~\ref{torpedo}.) The
planning in this CDR accounts for the cavern space needed to change-out and store
spent target torpedoes.

Neutrons flow out to the sleeve surrounding the target.  The sleeve, 1.2 meters in diameter 
and 1.3 meters long, contains a mixture of lithium and beryllium. 
Optimization of the sleeve shape is still ongoing, these dimensions are slightly
different from those mentioned in the Technical Facilities CDR.
The sleeve is surrounded by a 5 cm beryllium reflector, then by the steel and concrete
of the main shielding enclosure that is described in more detail in the next chapters.

\clearpage
\chapter{Transport and Installation}

This chapter covers the preparation of the site, including space and 
environmental requirements for the cyclotron and target caverns;  
modes and restrictions on transport of all the technical components 
from a point of origin in Japan to the KamLAND site, including offsite warehousing; 
rigging and other support for assembly of the cyclotron, 
MEBT (Medium Energy Beam Transport) line, and target.

\section{Cavern Preparation}

\subsection{Introduction}

The KamLAND detector is located in the cavern originally occupied by 
KamiokaNDE, a 3 kiloton water-Cherenkov detector.  
The cavern was built in 1982 in the Mozumi mine under Mount Ikenoyama 
in Gifu Prefecture, near the city of Toyama on the West side of Japan.  
After completion of modifications in the late 1980Õs, 
the current cavern has a cylindrical shape 20 meters in diameter 
and 20 meters high, with an additional dome on top that increases 
the maximum height to about 26 meters.  

\begin{figure}[t]
\centering
\includegraphics[width=5in]{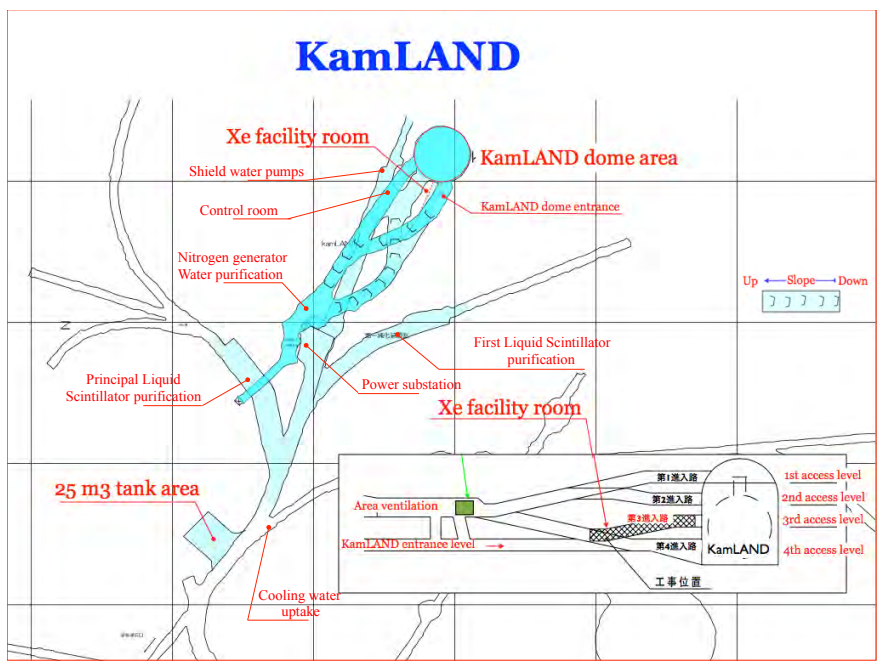}
\caption{{\footnotesize  Layout of drifts (passageways) in the KamLAND area.}
\label{KamMAP}}
\vspace{0.2in}
\end{figure}

Several drifts at different levels were excavated to facilitate the construction 
and removal of rock.  
The configuration of the detector cavern and the access drifts is shown in 
Figure~\ref{KamMAP}.  
Comparing the plan view with the elevation, one can see that the most 
intense blue marked KamLAND dome entrance corresponds to the 1st (highest) access level; 
the furthest to the left is the 2nd access level; 
the 3rd is marked ``Xe facility room'' almost directly below the 1st access; 
while the 4th, along the main entry way into the KamLAND area is the lightest color, 
in between the others.

As the experiments developed, different areas of these 
construction drifts were enlarged and refinished for support functions.  
A control room was built next to the 2nd access level; 
to the left of the lowest access are pumps for circulating the outer shield water layer; 
a water purification center and nitrogen generator area were placed in an 
enlarged area at the bend to the upper levels; and liquid scintillator purification stations 
were placed as shown.  The one marked ``first liquid scintillator purification'' will be 
used for loading the Zen microballoon with the proper mixture of scintillation fluid and $^{136}$Xe to balance 
the density of the balloon and ensure neutral buoyancy at the center of the KamLAND detector.  
Also shown is the area now used for the power station, installed capacity is 1.5 MW.  
At the bottom, close to the vertical plan inset is a water pumping station to draw cooling 
water from the underground stream that runs along the side of the principal access tunnel (drift).

Over time, as the principal experiment has evolved, the importance of different 
support areas also changed, with some being less needed for experimental support.  
For instance, the control room is still functional, but is no longer staffed, 
main control operations are now handled remotely from the support building 
in Higashimozumi, a nearby town about 5 km away.  
Similarly, the water purification plant is no longer used.  
The importance of the first stage scintillator purification area will decrease 
after the current xenon microballoon is filled and deployed.

\subsection{Cavern Construction for IsoDAR: General Issues}

\begin{figure}[t]
\centering
\includegraphics[width=5in]{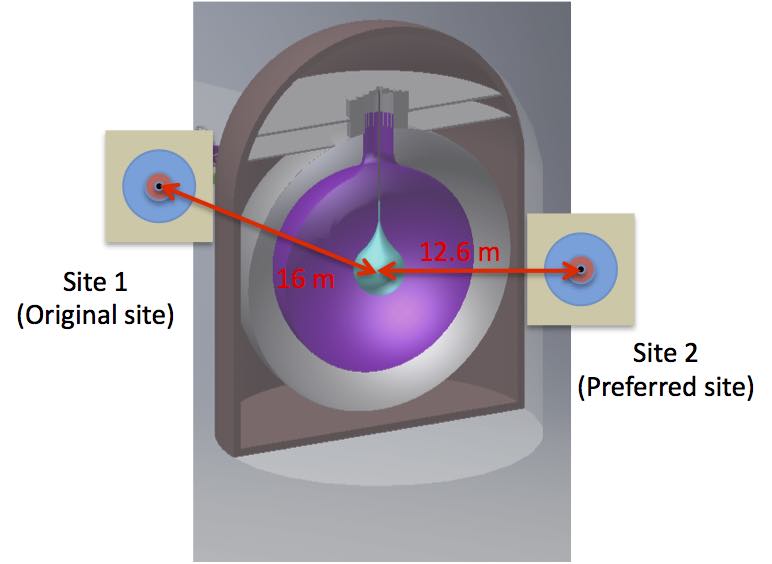}
\caption{{\footnotesize Vertical projection for the two sites evaluated for the placement
of IsoDAR.  Site 2, the preferred site, has the target right on the equatorial plane of KamLAND,
and hence can be almost 4 meters closer to the detector center.  This provides a
significant improvement in statistics and L/E.  Key to target system drawing:  tan = concrete,
blue = steel, red = Li-Be sleeve (antineutrino source for IsoDAR experiment), black = target.  Key to KamLAND:  brown = steel casing, void inside steel is filled with water; grey = mineral oil buffer and phototubes;  magenta = liquid scintillator; aqua = Zen microballoon filled with separated $^{136}$Xe and scintillator for presently ongoing neutrinoless double beta decay experiment}
\label{VertCompare}}
\vspace{0.2in}
\end{figure}

In this section, we provide information and considerations about
general issues, before providing location-specific scenarios.  Then,
in the following two sections, we  provide two locations scenarios,
called  Site 1 (the original site) and Site 2 (the preferred site).  
Site 2 places the target on the opposite side of the KamLAND
detector, in previously un-excavated rock.  Being unconstrained
by existing drifts, the target can be placed on the detector mid-plane.
 As shown in
Figure~\ref{VertCompare}, the two locations are at different levels
and lead to different distances to the center of the detector.
An important differentiating feature is that the original location
envisions two separate areas for the cyclotron and target, connected
by a long tunnel in which the much-more complex MEBT must be installed,
 while the preferred site
proposes a single cavern space for all of the equipment.
While both locations have pros and cons, operational and physics analyses
indicate Site 2 should be the preferred site.  Offsetting this is the increased
volume of rock that must be removed.
Nevertheless, both sites are feasible and we address general siting conditions for both
in this section.

\subsubsection{Gentle Rock Removal}

In either site for the experiment, we will see that caverns must be enlarged or excavated to house
the components.  In a hard-rock mine, such as this one, dynamite blasting is the usual
means employed for breaking up the rock.  However, 
blasting will not be possible close to the KamLAND detector because of the sensitive phototubes..  
Fortunately, ``gentle'' mining techniques are well known, are practical, and can be
employed to minimally-impact the sensitive electronic components of the detector.
One such technique is called ``expandable grout,'' in which appropriate holes are drilled, 
then filled with a cementaceous material that expands when it hardens, 
causing pressure on the rock surrounding the hole.   This leads to fracturing
in a way that does not cause the severe pressure waves that are generated 
when using dynamite.    

Once fractured, rock removal and disposal is an issue easily dealt with by the excavation contractor, 
almost sure to be the mining company managing the site.  Typical is to transport the rock to
previously excavated underground areas, so rock never reaches the surface, 
to be an environmentally-sensitive disposal issue.

\subsubsection{Cavern Size Considerations}

The IsoDAR layout can be thought of in three parts:  the cyclotron region,
the MEBT region and the target region.    

Specifications for each section are driven by space for technical components, 
shielding, assembly, and
maintenance requirements.

Staging areas and assembly methods are discussed below, in section \ref{AssySect},
from which we learn that in addition to the receiving and lay-down areas for technical components and shielding
blocks, space must also be provided for the hoist rails and support beams needed to move equipment
around in the cavern.  
Efficiency can be improved to minimize open staging areas if one carefully plans assembly steps so that
components are assembled and moved into position, freeing staging space for the next batch
of components.

Similarly, space and maintenance requirements can be facilitated by efficient utilization of staging space for
installation of mechanical and electrical systems once assembly of major components has been
completed.  In addition, one must keep
in mind that space must be available for service personnel to reach equipment requiring service.

The design and footprints of technical components are reeasonably well developed, so space requirements
are straightforward to specify.

The biggest demand for space will be shielding.  The requirements for stringent control
of rock activation, and preserving the low-background nature of the environment drive
the design of shielding, and as the amount of radiation from our cyclotron and target is so high,
the shielding will be very extensive.

Shielding design is also constrained by the need to access components for service.  One must provide removable
blocks, possibly accessed by the overhead cranes, or built on rails that allow sliding roof or wall areas out
to expose equipment inside.  All these design considerations must be taken into account in 
specifying cavern sizes.

The cyclotron is by far the most challenging component, with a mass of over 400 tons that
must be brought in (as we will see) in pieces no larger than 10 or 15 tons.  
A large staging area with heavy hoisting equipment will be needed to build up subsystems, 
and to assemble these into the cyclotron itself.
Support equipment for the cyclotron:  
RF amplifiers, power supplies, vacuum equipment and controls require a lot of space as well, but
as indicated could be installed in spaces used for assembling of the cyclotron after the machine
is installed in its place.

The electrical substation, providing 3.5 MW of power from the transmission lines brought to the site,
must also be allocated space.  This should be close to the cyclotron, as most of the power is
required there. The footprint for this substation can also be utilized as staging area for the assembly
of the cyclotron systems.

\subsubsection{Cavern Size Considerations -- Shielding}

(A more extensive discussion of shielding is contained in Chapter~\ref{radpro}.)

The stringent radiation-containment regulations in Japan mandate extremely careful calculations to design
the shielding for the cyclotron, MEBT and target.  We have developed sophisticated tools using the
MonteCarlo code GEANT4, extensively used by the nuclear- and high-energy-communities.  

We have determined that to prevent residual rock activation to levels above the 0.1 Bq/gm, we must
shield the target and sleeve with at least 1 meter of steel and 1 meter of concrete, the latter loaded with 
boron carbide aggregate to further enhance capture of slow neutrons.

Shielding requirement is driven by the ``source term" namely the amount of neutrons produced that must
be contained.  The source term for the target is well defined, as neutrons must be produced to generate
the neutrinos that drive the experiment.  

The source terms for the cyclotron and MEBT cannot be so easily defined, as neutrons arise from beam
loss, and this is a parameter we are attempting to minimize as much as possible.  However, the most
conservative approach is to say that, as we know how to contain 100\% beam loss (which is what
happens in the target), if we use the same shielding thickness everywhere we can be sure that
rock activation requirements will be met.  This is total overkill, as we know beam losses will be much less,
however we have not yet done the calculations for the cyclotron and beam-lines, so we cannot now 
specify a lesser amount of shielding that will satisfy the safety requirements.  The problem is
that beam loss locations and quantities cannot be completely defined, and the cyclotron is not
a homogeneous device.  The main iron pole and yoke are very massive, and provide a lot of 
shielding, however there are holes in this steel, for RF feedthroughs, vacuum lines, beam
injection and extraction, and diagnostics.  Neutrons can stream through these holes.  The most
efficient shield will have thick layers in the vicinity of these holes, and thinner layers in the
areas where cyclotron steel provides a good amount of self-shielding.  These calculations
have not been done yet, in part because the final detailed design of the cyclotron is not yet
complete.

A consequence of this conservative approach is that cavern sizes for the cyclotron will 
be substantially larger than what will ultimately be needed.  In a sense, this provides a buffer, in that
the optimization process will reduce the size requirements, so using the above specifications 
is in reality an upper limit on the cavern sizes required.

The MEBT has one area where neutron levels will be high, near the stripper foil.  This is close to the
extraction point, and it is planned to design the cyclotron shield to include this stripper area.  
The remainder of the MEBT is not expected to need much shielding, and can probably be instrumented
to cut beam off it a transient mis-steering of beam causes unacceptably high neutron levels.

\subsubsection{Cyclotron vault}

This is by far the largest area, dominated by the cyclotron itself.
This measures 6 meters across, is 3 meters high, and must allow space for RF stems, injection lines,
extraction lines, the afore-mentioned stripper station, as well as support mechanical and electrical 
systems.
In addition, the top half of the steel yoke must be raised by almost a meter,
to access the central region, RF dees, extractor and other parts that may need 
maintenance.   
As indicated above, additional space in the cavern is
needed for shielding.  Our initial conservative design is to provide 2 meters
of shielding around the entire cyclotron; one meter of steel, one of boron-loaded concrete.

To estimate the required cavern height, we start with 2 meters of shielding at the base, 0.5 meters between base
of shielding and bottom of cyclotron, 3 meters of cyclotron, 0.5 meters to the bottom
of the shielding blocks, 2 meters of top shielding and two meters for the crane, one
arrives at a required cavern height of 10 meters.

To
calculate the minimum width of the cavern, 
the shielded vault should have an inner dimension of about 8 meters, 
to allow 1 meter of access and clearance on all sides of the 6-meter cyclotron.
With 2 m of shielding, then, the width of the shielded enclosure, should be 12 meters.  
The end with 
the extraction port will be longer, to accommodate as well the shielding 
associated with the stripper, analysis magnet and dump for unstripped H$_2^+$ beam.  
This area can be tapered, and contoured to fit existing open spaces.

To estimate the size needed for the cyclotron cavern, we have a 12 meter
width of the outside of the shielding structure.  One side could be against
the rock face, but at least 2 meters would be needed on the other side for
personnel and service access.  The length of the shielding structure would
be the same 12 meters, but an additional 5 meters is needed to accommodate
the stripper shielding in the MEBT.  To accommodate this structure, the
vault must be at least 250 square meters. 

Considerations for the design and installation of the shielding structures are discussed later, in 
section \ref{ShieldAssy}.

The central region of the cyclotron, with the injection RFQ and ion source, will protrude 
below the midplane, and probably extend at least one meter below the bottom steel. 
However, this is a low-radiation area with few neutrons, so having a pathway through 
the 2 meters of bottom shielding may provide adequate access to the source and 
RFQ for maintenance. There should be no need to provide a below-ground
pit in either case.

Service equipment for the cyclotron, the RF amplifiers, power supplies and controls,
will require floor area of about 220  square meters, to place racks and allow
space to access them for service and maintenance.  

The 3.5 MW electrical substation will require about 100 square meters, again
allowing adequate space for insulation and isolation, and for service and maintenance
tasks.  

\begin{figure}[t]
\centering
\includegraphics[width=5in]{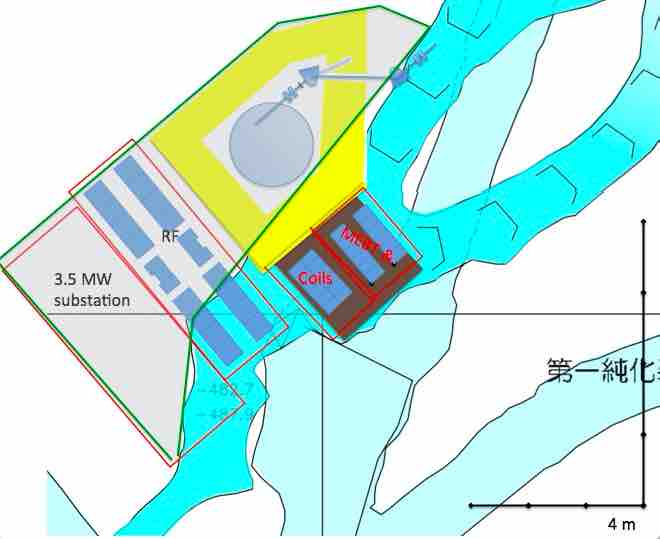}
\caption{{\footnotesize Area required to install the cyclotron, electrical substation
and power systems for the cyclotron and MEBT.  New excavation is required, 
about 200 square meters with 10-meter-high ceiling over the cyclotron and RF
supplies (to retract the roof for cyclotron maintenance), 
and 100 square meters with ceiling 4 meters high over
the electrical substation.  The grey area shows the location of the new excavation.}
\label{Installed-Cycl}}
\vspace{0.2in}
\end{figure}

The total floor space for these three systems is 470 square meters.

Both of the electronic support areas do not need the 10 meter ceiling height,
however reach of the crane rails would indicate a preference for the same height
specified for the cyclotron shielding, at least over a portion of the 
electronics space.  The volume of the cavern would then be approximately 5200 
cubic meters.

A possible layout of the cyclotron vault can be seen in Figure~\ref{Installed-Cycl}

Again, these are generous estimates, without benefit of careful layout and design, so
should really be viewed as upper limits of the actual space needed.

\subsubsection{Target Vault}

\begin{figure}[t]
\centering
\includegraphics[width=5in]{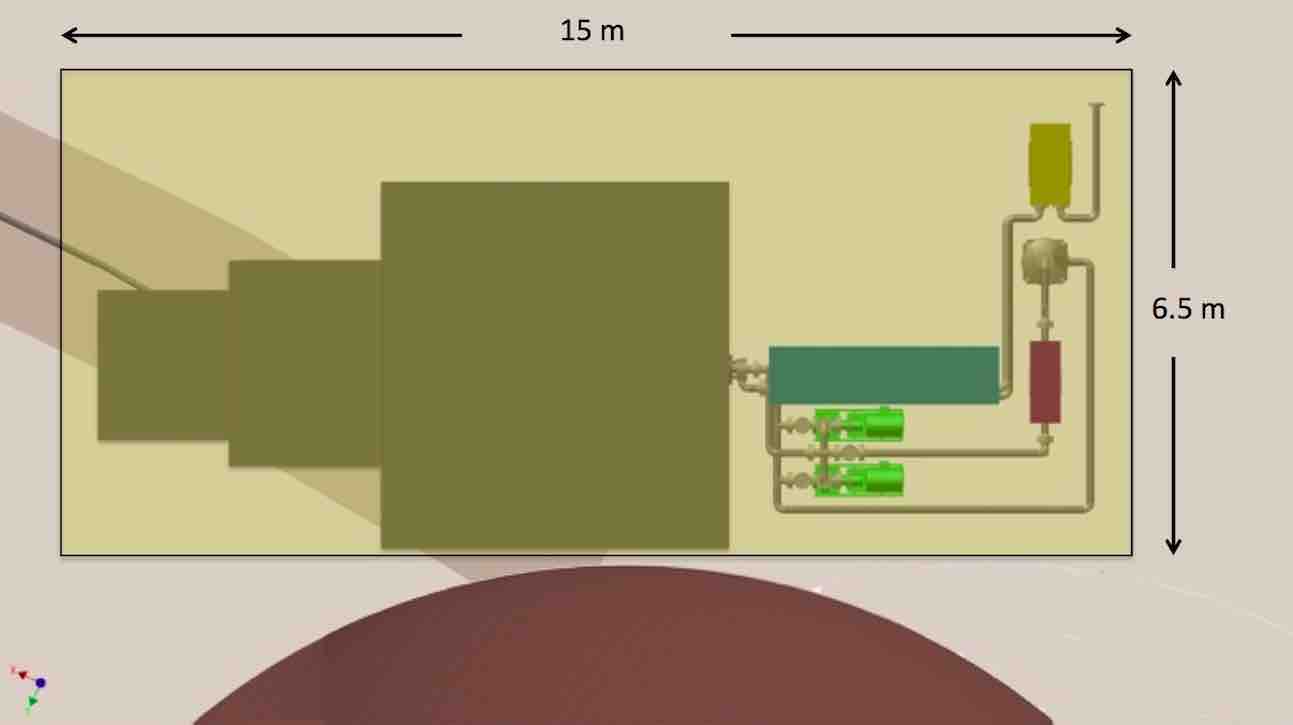}
\caption{{\footnotesize  Plan view of target showing shielding (5 meter width and height) that provides
adequate neutron protection.  The dark-green-colored box shown downstream is 
a casket to house a spent target.  Target changing must be accomplished
by remote handling.  The vault needed to house the target systems,
approximately 15 meters by 6.5 meters, with a 6 meter ceiling height,
 is superposed. }
\label{TgtPlanView}}
\vspace{0.2in}
\end{figure}

The overall dimensions of the target shielding structure is 5 meters wide
and 5 meters high, and 8 meters long. However, as seen in 
Figure~\ref{TgtPlanView}, space is needed for the water cooling
system for the target, as well as for the caskets needed to exchange
the target torpedos.  Space for staging, assembly, and personnel
access leads to a vault size of 15 meters by 6.5 meters.  
As the shielding is 5 meters off the floor, at least 6 meters is needed for
the ceiling height.  But another meter should be added for crane access.

Total volume of the target cavern is then 680 cubic meters.

\subsubsection{MEBT space}

The cavern requirements for the MEBT depend upon the choice of
site.  We consider the impact of the MEBT on the caverns in the
individual cases below.

\subsubsection{Other Considerations}

Two other points to note about the cavern design: First,
the required spans that are planned are well within typical cavern sizes at the
Kamioka mine--none involve unusual engineering issues.  Second, with
safety in mind, designs must include a route of secondary egress.

For environmental control, the cavern should be lined with the same heavy
polyethylene material currently used in other KamLAND support areas where
clean conditions must be maintained.  See Figure~\ref{polyLiner},
which is discussed later in this CDR.

\subsection{Site 1, the Original Site Location}

This site makes use of existing spaces for the cyclotron and the
target, with a drift (passageway) that can accommodate the MEBT.
Equipment in the existing spaces would need to be removed, and in some
cases, relocated.  This site was first put forward with the idea that we could avoid
cavern construction, which we initially believed would require
blasting.  However, upon careful analysis it was realized that the
spaces available were too small to accommodate the equipment, so in 
any event substantial enlargement of the caverns is required.

The problem of blasting has been mitigated though by the techniques for gentle rock
removal.   This technique is expensive, though, so existing caverns still represent 
an economic benefit.

On the other hand, as discussed below, installation of the experiment in existing 
areas that are accessed at least once or twice a day by KamLAND personnel
in order to check equipment in the KamLAND dome, presents a great inconvenience
to both IsoDAR and KamLAND.
While the
detector is being accessed, the cyclotron RF must be off, so there will
be an interruption in the beam.  Restoring stable operation is time-consuming 
and inefficient.

\subsubsection{Cyclotron Region}

\begin{figure}[t]
\centering
\includegraphics[width=5in]{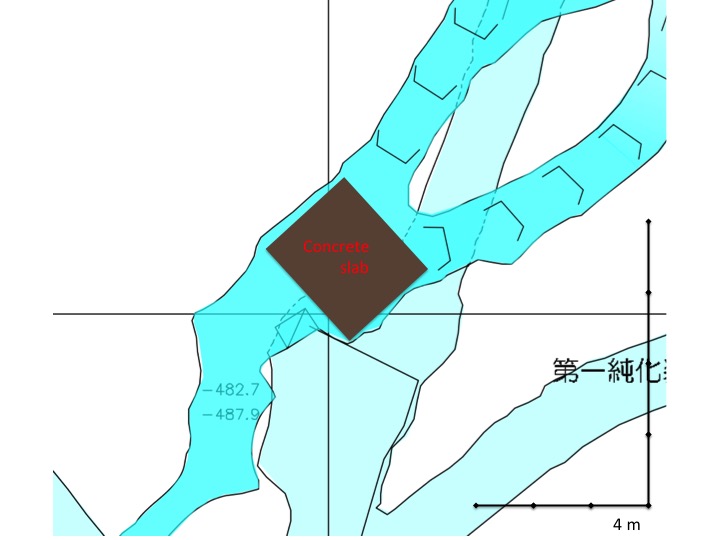}
\caption{{\footnotesize  Suggested site for the cyclotron:  the area currently housing water-purification and
nitrogen gas systems.  The concrete slab measures 8.7 by 7.1 meters.
 }
\label{Existing-Pad}}
\vspace{0.2in}
\end{figure}

The original site places the cyclotron region components in the area of the currently unused 
water treatment area, and the nitrogen gas generator, shown in
Figure~\ref{Existing-Pad}
The equipment currently in this area would need to be removed, and the
cavern enlarged to accommodate the cyclotron, its shielding, and the
power systems.

The space has a high ceiling, which is advantageous, but as we saw in Figure~\ref{Installed-Cycl}
the area required for the cyclotron and support equipment is considerably larger.
The area to be excavated is shown in the shaded region of Figure~\ref{Installed-Cycl}.  
Approximately 2000 cubic meters of existing space could be incorporated into
the cyclotron vault, so the
 net volume of new excavation in the cyclotron region would be about 3200 cubic meters.

\subsubsection{Target Region}

\begin{figure}[t]
\centering
\includegraphics[width=5in]{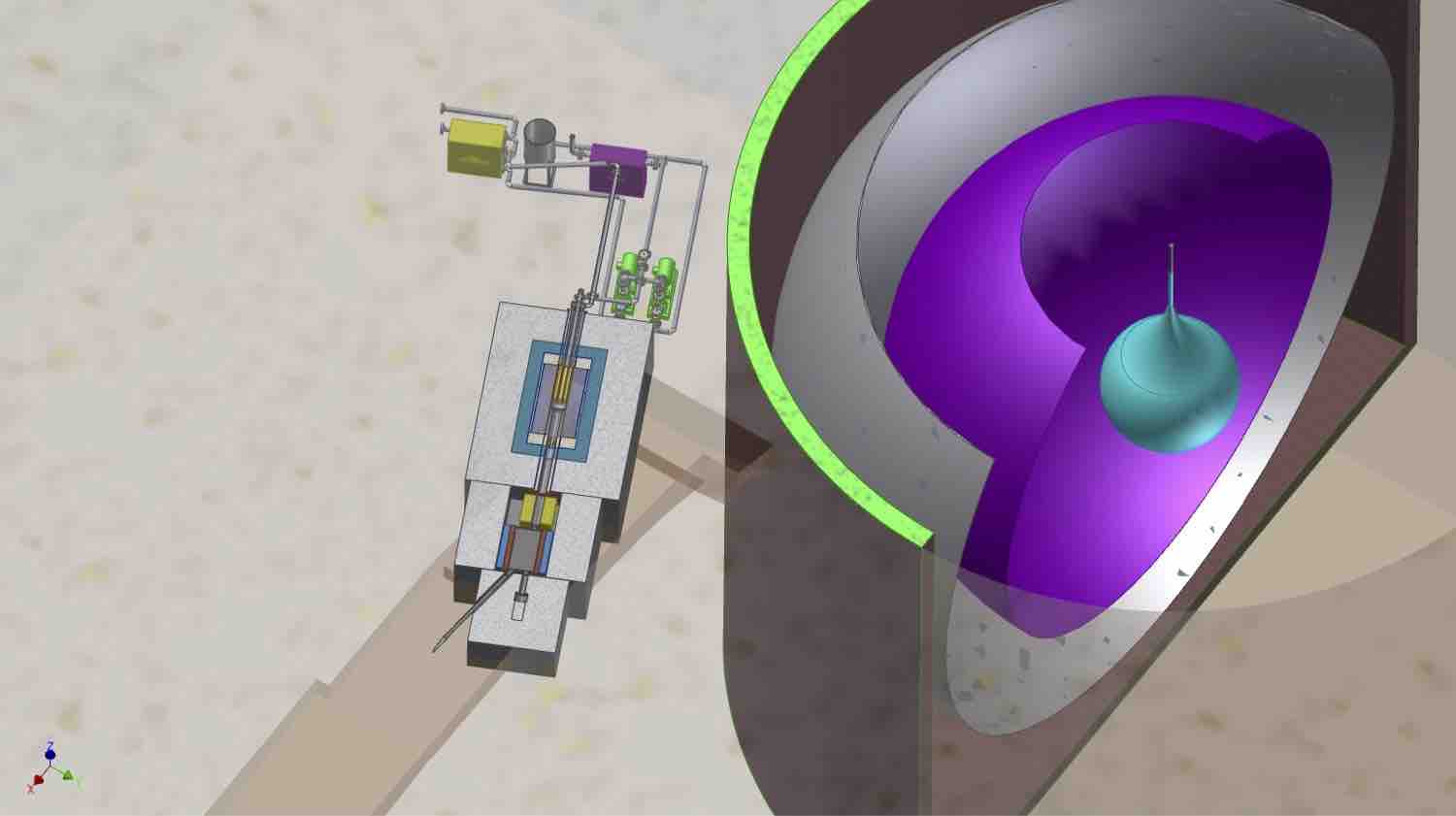}
\caption{{\footnotesize  Target next to KamLAND.  30\degree bend in beam line
directs backstreaming neutrons into dump.  Shielding thickness has been increased
more than shown.  Primary water cooling components are shown. Downstream space is
required for target changing.  The light background is rock, existing drift is shown in darker tan. 
Enlargement of the area will be needed to accommodate the target and support systems.}
\label{tgt3d}}
\vspace{0.2in}
\end{figure}

The existing space is a control room that would need to be relocated to an new
site.  This leaves a cavern with a footprint shown on 
Figure~\ref{tgt3d} by the brown area on the left, and the existing
rock is shown in mottled tan.   This figure
shows the target assembly, including the water
circuit system, indicating how the system would be oriented with
respect to the detector on the right.  
Updated shielding studies add more mass to the target assembly;
an updated
plan view is provided in Figure~\ref{TgtPlanView}.  As described above,
a 30\degree bend in beam line
directs backstreaming neutrons into dump.
The area downstream of the target requires additional space to
accommodate target change-out.    
The target-changing ``casket'' and equipment for the primary coolant
loop and heat exchangers set the length required for the 
vault needed to house the target assembly.  
Figure~\ref{TgtPlanView}
also shows, in light color, the size of the vault.  
%
%
%
 %
 
The required cavern dimensions: 15 meters long, 6.5 meters wide and
 7 meters high contains a volume of 680 cubic meters. 
The approximate size of the existing drift is 4 meters wide by 3 meters high, and
 the overlap length with the required cavern is approximately 9 meters.  
 The volume of existing empty space is 110 cubic meters.
 Thus the volume of new excavation is about 570 cubic meters.
  
The design provides a 
 secondary escape route through 
a drift almost directly under the upper right corner of this cavern.
A vertical rise will be drilled to this drift, approximately 10 meters below, 
 adding approximately 20 cubic meters of extra excavation.

 It would be convenient and straightforward to plan on storing spent
 targets (and fresh replacement targets) in bore holes drilled horizontally
 into the wall away from the KamLAND vault.  
 These could be accessed by a remote handling arm.  
 These boreholes should have a steel pipe lining, the same diameter
 as the vacuum pipe in the target assembly, and should be about 
 3 meters deep, allowing for a steel plug to provide shielding from gamma
 radiation from the spent target.  
 Each bore hole would require about 2 cubic meters of excavation.  
 If six are provided, one should plan for an extra 12 cubic meters of rock removal.

In total, for the target area, about 600 cubic meters must be
removed.  
 
\subsubsection{MEBT Beam Transport Line Region}

\begin{figure}[t]
\centering
\includegraphics[width=5in]{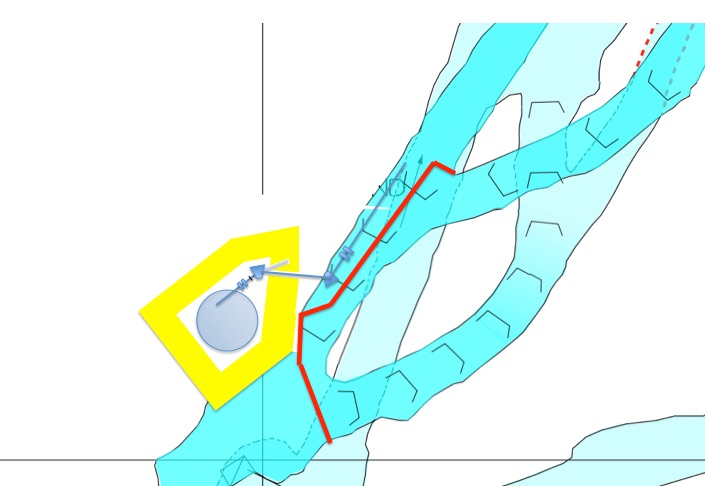}
\caption{{\footnotesize The MEBT (Medium Energy Beam Transport) line would run along 
the left wall (looking towards KamLAND).  A proposed full-height wall, shown in red, divides the passageway into
two sections, providing some shielding, but also is part of a fire-life-safety system.  Personnel in the KamLAND dome area can evacuate safely if there is a fire in the electrical areas.  Personnel in the IsoDAR area can evacuate through a fire door in this wall by the electrical systems, out the escape chute in the target vault, or through a new escape exit in the back of the electrical area connecting to the lower level.}
\label{MEBT-wall}}
\vspace{0.2in}
\end{figure}

The MEBT is to be installed in an existing drift that connects the
target and cyclotron regions.
The passageway is adequate for this line, requiring little or no added excavation.
However, vertical slopes in the drift (up then down)
require additional vertical steering magnets in the transport line, increasing its
complexity and allowing more opportunities for beam loss. 
There is room to place shielding, 
up to a meter, along the beam path, which is probably more than is be needed, however
no specific calculations have been done yet.
A full-height wall will divide this passageway into two paths, one for personnel
to access the KamLAND dome, the other for the beam line.  The width of the drift is about 5 meters,
each new passageway would be about 2 meters wide.  
Lining this concrete wall with lead or a steel plate, would provide shielding from residual gamma
radiation from the beamline components.  It could also provide environmental separation that might be
important when evaluating life-safety factors, and providing separate, isolated evacuation routes in the event of 
a fire.
Figure~\ref{MEBT-wall} 
shows schematically how the MEBT and separating wall might be placed.

\subsection{Site 2, Preferred Site for IsoDAR}

\begin{figure}[t]
\centering
\includegraphics[width=5in]{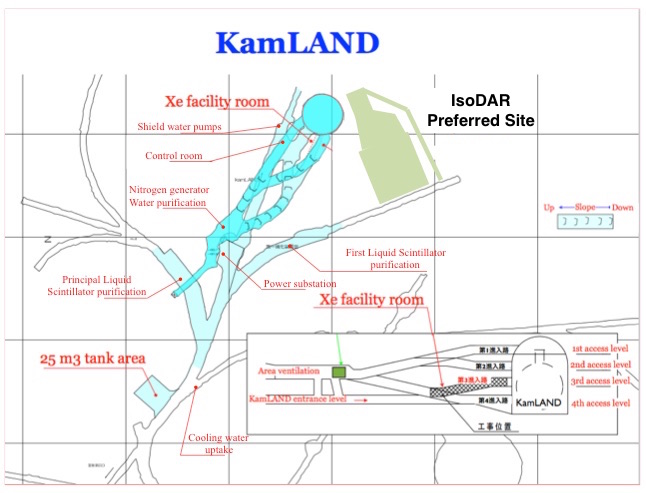}
\caption{{\footnotesize Suggested new location for IsoDAR on opposite side of KamLAND.  Access drift is from existing ``First scintillator purification" area. }
\label{new-site}}
\vspace{0.2in}
\end{figure}

The proposed alternate site offers many advantages.  It is located on 
the opposite side of the KamLAND detector.
This site requires excavation of a new cavern, removing substantially
more rock.   However, on balance, there are a number of advantages to
this space, as we discuss below, especially noteworthy is the ability to place the target
right at the mid-plane of the detector, increasing the detector solid angle.

\subsubsection{The Site 2 Cavern} 

Figure~\ref{new-site}
shows the new cavern, with the neutrino source approximately 180\degree from the 
original location.  One single larger cavern would house all the components, with some space
savings by being able to utilize the same space, and rigging equipment, for
assembly of both the target and cyclotron.  In addition, the MEBT would be more compact,
would have fewer elements (in particular, eliminating the vertical bends), 
with a high likelihood of less beam loss.

\begin{figure}[t]
\centering
\includegraphics[width=5in]{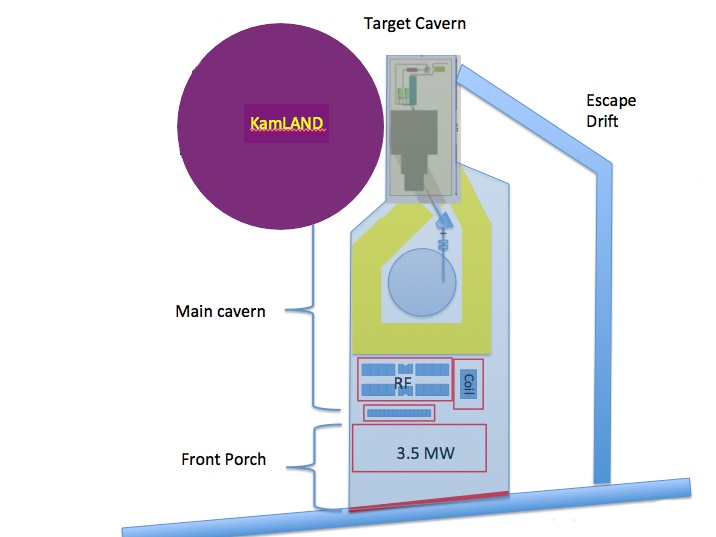}
\caption{{\footnotesize Layout of equipment in the new cavern.  Escape drift connects the upper 
corner of the target cavern with the main access drift.  After equipment is delivered and installed, a fire wall (red line) with an access fire door provides a fire barrier for the new area. Total new excavation would be around 7200 cubic meters, including secondary egress and enlarged access route from the present scintillator purification area. }
\label{Opt2detail}}
\vspace{0.2in}
\end{figure}

Placement of the cyclotron, target and support electrical equipment is shown in 
Figure~\ref{Opt2detail}.
Cavern ceiling heights mirror those described above for the original target and cyclotron areas.

Access to this new cavern would be from the area marked as ``First scintillator purification"
which, as indicated earlier, will be less utilized once the Zen xenon balloon has been filled.
The required 4.5 meter elevation change can be accommodated by sloping the access drift
beyond the end of the present purification area.     A large parking
area is located very close to this entrance.

Secondary egress, for personnel safety, could be provided from the far corner by the target, 
with a drift extending back to the access drift, a length of about 60 meters.

The volume of the new cavern can be estimated by adding the sizes of the target and cyclotron vaults, 
the space required for a compact MEBT, and the footprints for the new 3.5 MW 
substation and accelerator power supplies. 
Vault heights would be the same as the corresponding areas of Option 1:  
7 meters in the target area, 10 meters in the cyclotron vault, 
and 4 meters in the substation area, labeled ``front porch."
 The escape route adds an estimated 300 cubic meters, and the enlarged
 access route another 200.
The total volume of new excavation is about 7200 cubic meters.

\subsubsection{Cost-Benefit Analysis of the New Site}

Comparing the rock removal between the two sites, 3800 vs 7200, we see
that an additional estimated 3400 cubic meters would need to be removed.
The added cost is substantial.  However, there are considerable
programmatic and practical advantages that should be considered to offset this additional
cost.  Quantitative estimates can actually be made, these programmatic values will be calculated
 for presentation in the PDR.
 
 In addition, it should be reiterated that the cavern size requirements are rough estimates,
 careful engineering studies are required to reach more accurate designs for placement
 of equipment and cavern sizes, especially after more detailed shielding calculations are performed.
 
 But rather than pure cost analyses, probably of more value are the programmatic advantages of the
 second site.
 
 As mentioned earlier,  the elevation of the target can be placed exactly on the 
equatorial mid-plane of the detector balloon.  
From Figure~\ref{VertCompare} one can see that the target in the 
Control-Room drift (original site) is located at about the top of 
the balloon containing the liquid scintillator, almost 7 meters above 
the mid-plane of the detector.  The mid-plane location of the preferred site allows
 placing the neutrino source almost 4 meters closer ($\sim$12.6 meters instead of $\sim$16 from the 
detector center), with substantial improvement in the physics reach of
the experiment.

If the collaboration decides to trade the additional rate for reduced
power in the beam, this could have considerable cost advantages for
both installation and running.   Less power and water would need to be
installed, reducing the cost of initial utilities and operations.
There may also be savings in improved control of environmental factors, such as air circulation and ground water management. 

Technical risks are also reduced, as the beam in the cyclotron would be reduced
 from 10 mA (of stripped protons) to 6 mA.  This would enable easier commissioning, and
 provide a very welcome performance margin for the accelerator team to shoot for.

Another advantage of the second site
 related to interfacing with KamLAND would be the need to relocate or remove much less equipment.
 
 A major benefit, though, is the complete independence of operation of KamLAND
 and IsoDAR with the second site.  Access to the KamLAND dome could probably
 be allowed even when beam is being delivered to the IsoDAR target, as opposed
 to having to shut the IsoDAR system off, survey for radiation safety before allowing
 personnel to access the KamLAND area, then secure it again and bring the IsoDAR beam
 back up.  This is terribly wasteful in running time, and is a great inconvenience for operations
 of both experiments.

\section{Transport of Equipment to the KamLAND site}

\subsection{Transportation to the Kamioka mine}

Transportation to the site, storage, rigging and assembly of the IsoDAR components
will be a complex task, requiring coordination
with KamLAND and Mining Company staff.  
Equipment will be arriving from two principal sources:  overseas shipping, 
and local (Japanese) sources.  
The principal shipping point will be the Toyama harbor, located approximately
100 km from the Kamioka mine.  
Local sources are locations throughout Japan.

A warehouse will be used for temporary storage of materials.  Ideally, 
the location would be approximately midway between 
the Toyama harbor and the Kamioka mine.  
The warehouse can be used for local and overseas equipment as necessary.  
The shielding material, concrete and steel, may be stored in an outdoor facility, 
which could save on storage costs and simplify handling of the bulkier shielding blocks.

\begin{figure}[t]
\centering
\includegraphics[width=6.5in]{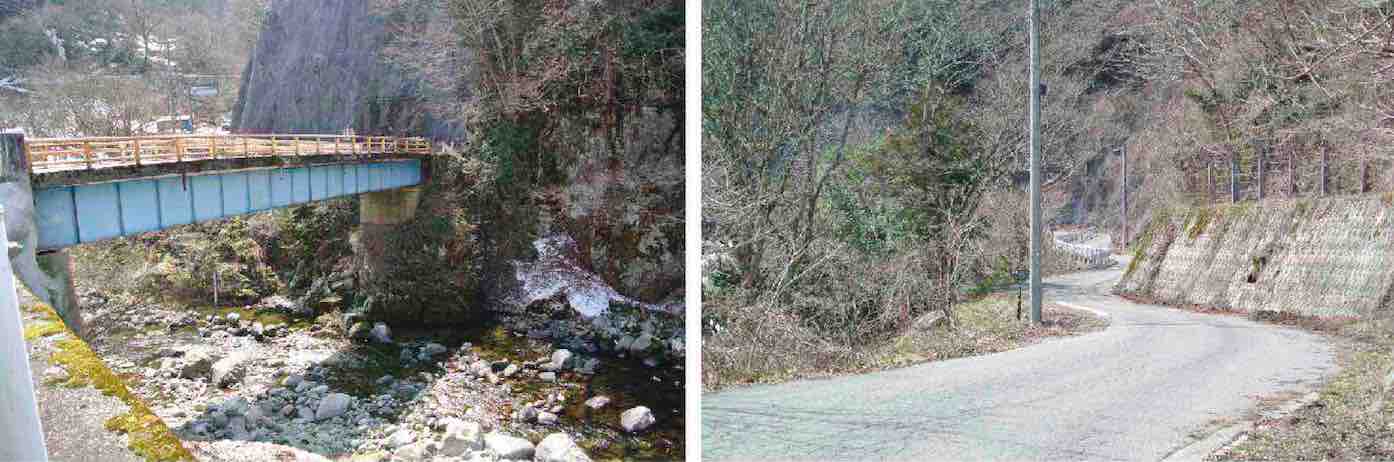}
\caption{{\footnotesize Photos of access road, and bridge.  Shipment
    along this route must account for the 
9-ton gross weight limit of the bridge and narrowness of the roadway. }
\label{road-bridge}}
\vspace{0.2in}
\end{figure}

\begin{figure}[h]
\centering
\includegraphics[width=6.5in]{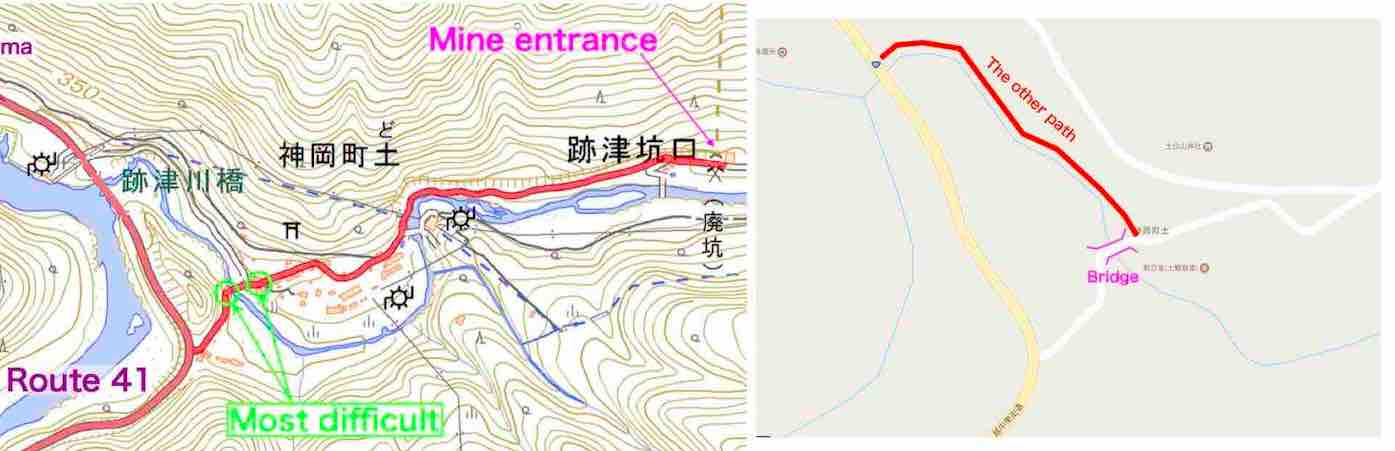}
\caption{{\footnotesize Maps of the roadways between the main highway, Route 41,
and the Atotsu entrance of the mine.  The principal route, depicted in Figure~\ref{road-bridge},
is shown on the left.  The alternate road, which bypasses the Atotsu bridge, is seen on the right. }
\label{route-maps}}
\vspace{0.2in}
\end{figure}

Shipping from the warehouse to the mine will take special consideration.  
The main roadway, Route 41, provides good access up to the branch-points
to the mine.  
However, it is a two-lane road, for which transportation of oversized loads would be
a problem.  For instance, the cyclotron coil, 5 meters in diameter, is too wide to load
flat on a truck, and would be too high if mounted vertically to pass under bridges.
Similarly, the vacuum chamber for the cyclotron would be too large to transport.
Both of these will need to be segmented and assembled underground.
But, between Route 41 and the mine entrance, problems become even more severe.
  The most commonly used, and easiest access to the mine is a narrow
local road with sharp
curves, about 1.5 km long.  
Figure~\ref{road-bridge}
pictures a portion of this roadway.  The primary limit on this roadway
is a bridge over the Atotsu tributary
with a weight limit (total: truck plus load) of 9 tons.  This is also seen in the same figure.
Alternatively, as seen on 
Figure~\ref{route-maps},
 there is a road from Route 41 which bypasses
the bridge.  This road, accessed by permit only, is also narrow and windy, but does not
have the severe weight limit imposed by the presently-used bridge.
The roadbed of this bypass is not in good condition either.
In view of these limitations, we are planning on loads no heavier
than 10 to 15 tons.
We also assume that a flatbed truck with a wheel-base of
about 9 meters will be the longest truck used, in order to handle the curves.

The plan for assembly of the cyclotron and target given these
restrictions has been published in the Technical Facility CDR.

The longest pieces to be transported will be the shielding roof
blocks for the cyclotron. These have a total length
of 10 meters.  This load would extend about 1 meter beyond the rear end
of the truck, but with proper flagging and permits it should be possible
to transport these long blocks.


	

\subsection{Transportation within the Kamioka mine}

Transport within the mine must meet the mine safety regulations 
(e.g. fire suppression, emission levels).  
In addition, the height of the truck, plus load, is restricted by the height of
the top of the access-route drifts (passageways) into the KamLAND area.
If the shipping trucks bringing loads along Rt 41 can meet the mine standards, then
most materials can be driven directly into the mine and offloaded at the entrance to the
KamLAND area.

 Because of the height of the truck bed, some pieces will be too large to be transported directly
 into the mine.  In such cases, offloading must take place at the mine entrance.  
 Such pieces will be loaded on specialized low carts and pulled in with tractors.  
 The mining company has this equipment.
 
 Transport from the KamLAND entrance to the staging areas will depend on which site
 is developed for the IsoDAR experiment.

For Site 1, access will be further restricted by drift width, ceiling height, sharp turns, and ramp slopes.
The mining company crews are experienced with transport of bulky and heavy pieces, 
using, for instance a forklift or overhead chain hoists at the offload point at the base of the KamLAND detector. 
Loads are rigged onto low-profile hand-carts, and winches are utilized to pull these loads
 up the inclines.  (See Figure~\ref{mine-equipment}.)
 A heavy-duty electric jack may also be necessary for maneuvering components around the sharp corners
  to reach
 the staging areas.
 
 The second site may be easier to access.  Elevation differences are less, and passageways will probably
 be enlarged with the specific objective in mind for less complications for bringing in the equipment.

\section{Support for Assembly and Installation\label{AssySect}}

\subsection{Hoisting}

\begin{figure}[t]
\centering
\includegraphics[width=6.5in]{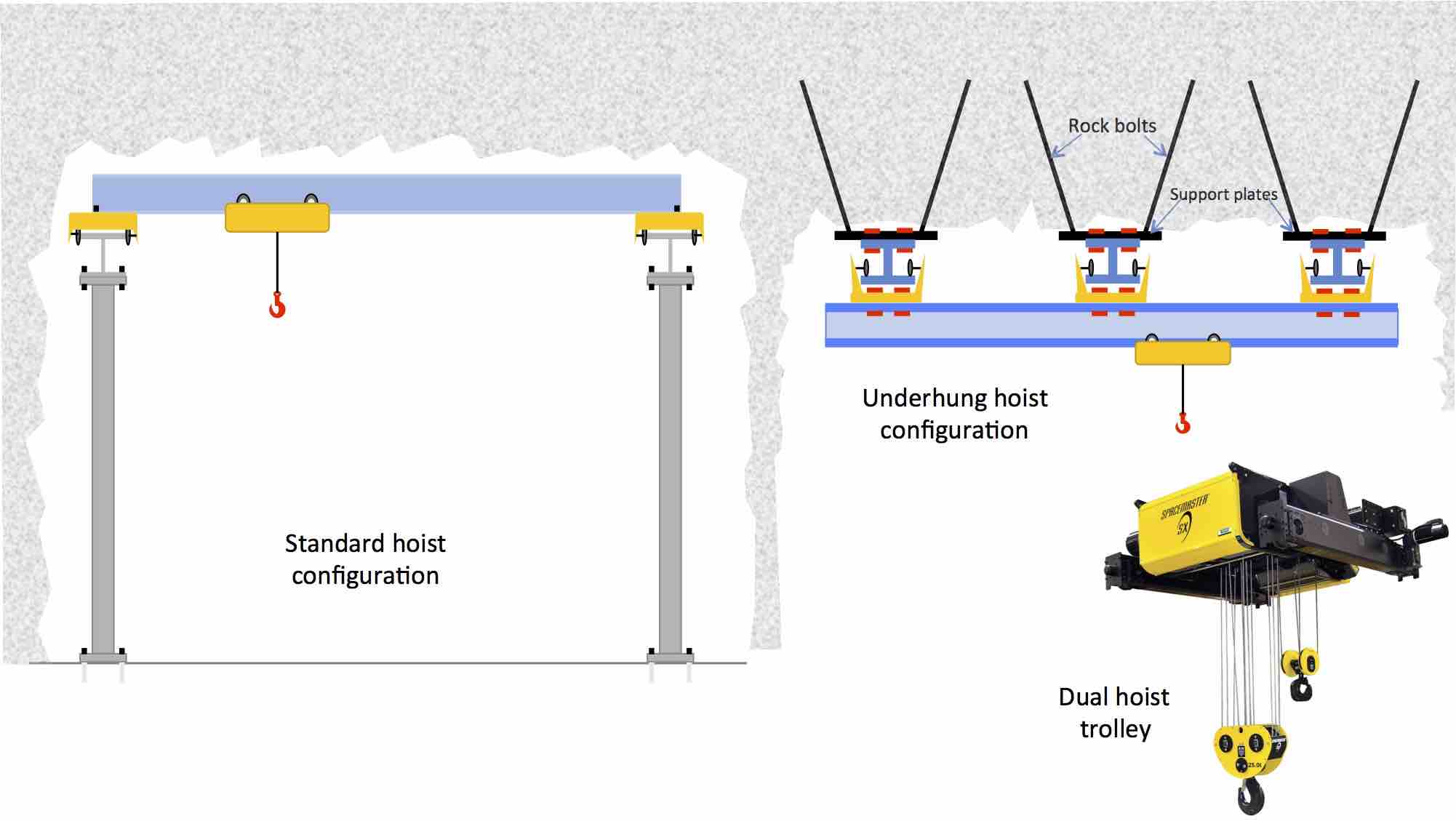}
\caption{{\footnotesize Overhead bridge crane configurations.  On left side is the 
standard configuration, with rails supported on posts from the floor of the cavern.
In the ``underhung'' configuration (right side), the crane rails are supported
by several  plates, about 1 meter on a side, each plate secured to the rock ceiling by
at least four rock bolts.  The trolley shown has two hooks, the main one is 25 tons, 
the secondary one 10 tons.  The larger capacity hook will be needed as heavy
cyclotron pieces are combined into sub-assemblies.}
\label{rigging}}
\vspace{0.2in}
\end{figure}

It is envisioned that overhead hoist systems will be installed in the IsoDAR caverns.
One should cover both the staging and assembly areas at the cyclotron location and 
probably a second system at the target location.
These will be used for offloading components brought in on carts and for 
assembly of cyclotron, target, and shielding.  
The maximum weight of individual pieces brought in to either vault is 10-15 tons.  
However, the hoist in the cyclotron vault will need a larger capacity for manipulating 
cyclotron sub-assemblies.  
It will probably be possible to assemble the target system, and stack shielding
around it without having to lift pieces heavier than 10 tons, so one might possibly
not require an overhead crane in this area, as long as there is sufficient room
to maneuver a heavy fork lift or cherry picker inside the target cavern.

The cyclotron hoist will have to provide coverage over an area that is 12 m wide, 
and at least 20 meters long; while a hoist in the target area, should it be needed,
 would span an area approximately 6x15 meters.  
 In both vaults, it will be important to minimize the headroom needed for hoisting to avoid excavating more rock than needed.  
 Two options can be considered for hoist support: standard and underhung as shown in
 Figure~\ref{rigging}.

Standard overhead hoists systems place the hoist beam (bridge) as close to the 
ceiling of the vault as possible.  
The hoist beam is supported at each end by runway beams.  
The ends of the runway beams are supported by columns anchored to the vault floor. 
 In this case, the height (size) of the hoist beam needed is proportional to its span length.
For an underhung hoist, the runway beams are attached to plates firmly anchored to the 
 ceiling of the vault with numerous rock bolts.  Such a configuration, supported
 by proper engineering calculations, should be
 capable of holding the planned loads \cite{SURF}.  
 The hoist beam could then spread its load over more than two runway beams, which is not
 possible in the standard configuration.  
 This makes it possible to increase the span of the hoist beam without increasing its height.
 
Both systems work with cable or chain hoists, but cable hoists require less headroom than chain hoists. 
  Overhead cable hoists can also be equipped with a secondary line to assist in flipping pieces.
  
\begin{figure}[t]
\centering
\includegraphics[width=6.5in]{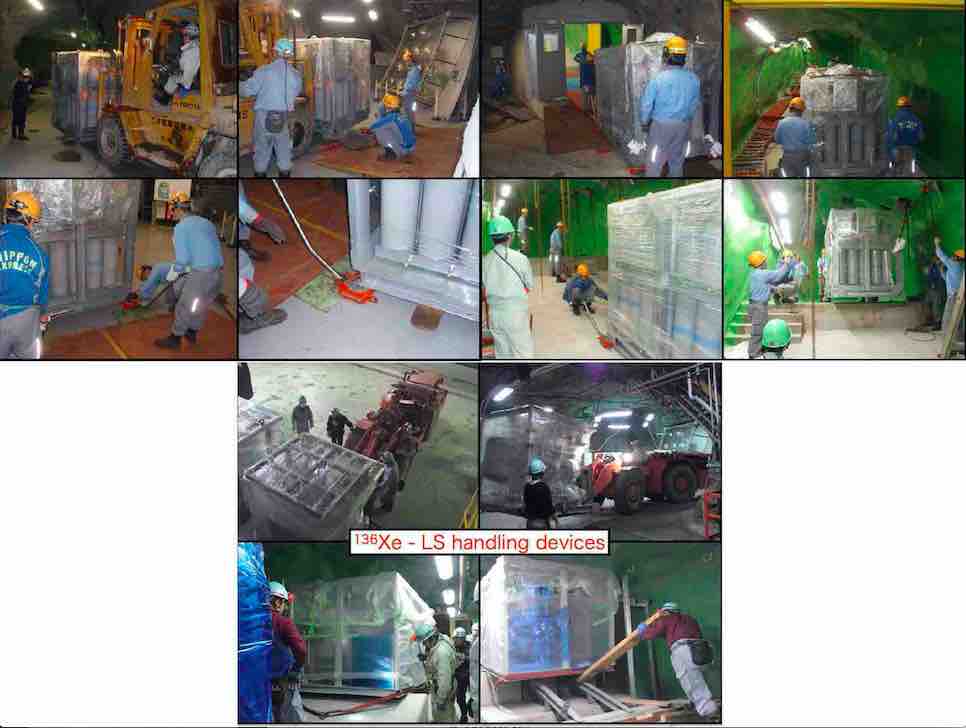}
\caption{{\footnotesize Photos from installation of the xenon systems for the Zen 
experiment.  Note the diversity of rigging equipment employed, from forklifts to
hydraulic lifters.  In particular, note the use of ceiling hooks (second row right side).}
\label{mine-equipment}}
\vspace{0.2in}
\end{figure}

Assembly of additional components (e.g. MEBT, electrical systems) may require other lifting methods.  
  Forklifts, carts and chain hoists can be used for this process, as KamLAND has done in the past.  
  Figure~\ref{mine-equipment}
  shows examples of recent rigging experiences at KamLAND.
  Such equipment will also be employed when the primary hoist is not sufficient for flipping or lifting large pieces and sub-assemblies.

\subsection{Shielding Assembly\label{ShieldAssy}}

\subsubsection{Introduction}

As already stated, the shielding configuration for the cyclotron has not
yet been optimized.  To demonstrate a solution for safely deploying
and operating the cyclotron, we have adopted the extremely conservative
approach of applying the shielding thickness calculated for the target,
and completely enclosing the cyclotron with this shield.

This is overly conservative, since the number of neutrons produced in the
vicinity of the cyclotron are orders of magnitude lower than at the target,
and in addition the 400 tons of steel of the cyclotron magnet provide
a very large amount of self-shielding.  

However, until the calculations can be performed, the safest approach
is to adopt this very conservative shielding plan for the cyclotron.
When the calculations are done, it is almost surely guaranteed that
the size and costs for the cyclotron enclosure will be substantially reduced.

\subsubsection{Assembly of Cyclotron Shielding}

All shielding will be shipped in 10-ton blocks.  The steel shielding above the cyclotron will need to span 10 m.  
The top blocks could then be 1 m wide and 12.5 cm tall (using a density of 8 kg/m$^3$ for steel).
  It will take 64 such blocks to enclose the top of the cyclotron with 1 m of steel shielding.  
  Concrete blocks (density 2.4 kg/m$^3$) could be made 1.25 m wide and 33 cm tall to span 10 m.  
  24 concrete blocks could then be used for 1 m of shielding above the cyclotron. 
   Additional blocks around the perimeter are needed to meet shielding requirements in all directions. 
    For 10 ton pieces, this will take another 16 steel blocks and 12 concrete blocks.  
    The bottom shielding has the same dimensions as the top.  
    The side shielding can consist of 4 steel walls 9x1x6 m each, 11x1x8 for concrete. 
     That is 172 steel blocks and 88 concrete for wall shielding.  
     Total counts are 332 pieces of steel and 160 concrete, approximately 500 truck loads for cyclotron shielding.
     
     \begin{figure}[t]
\centering
\includegraphics[width=6.5in]{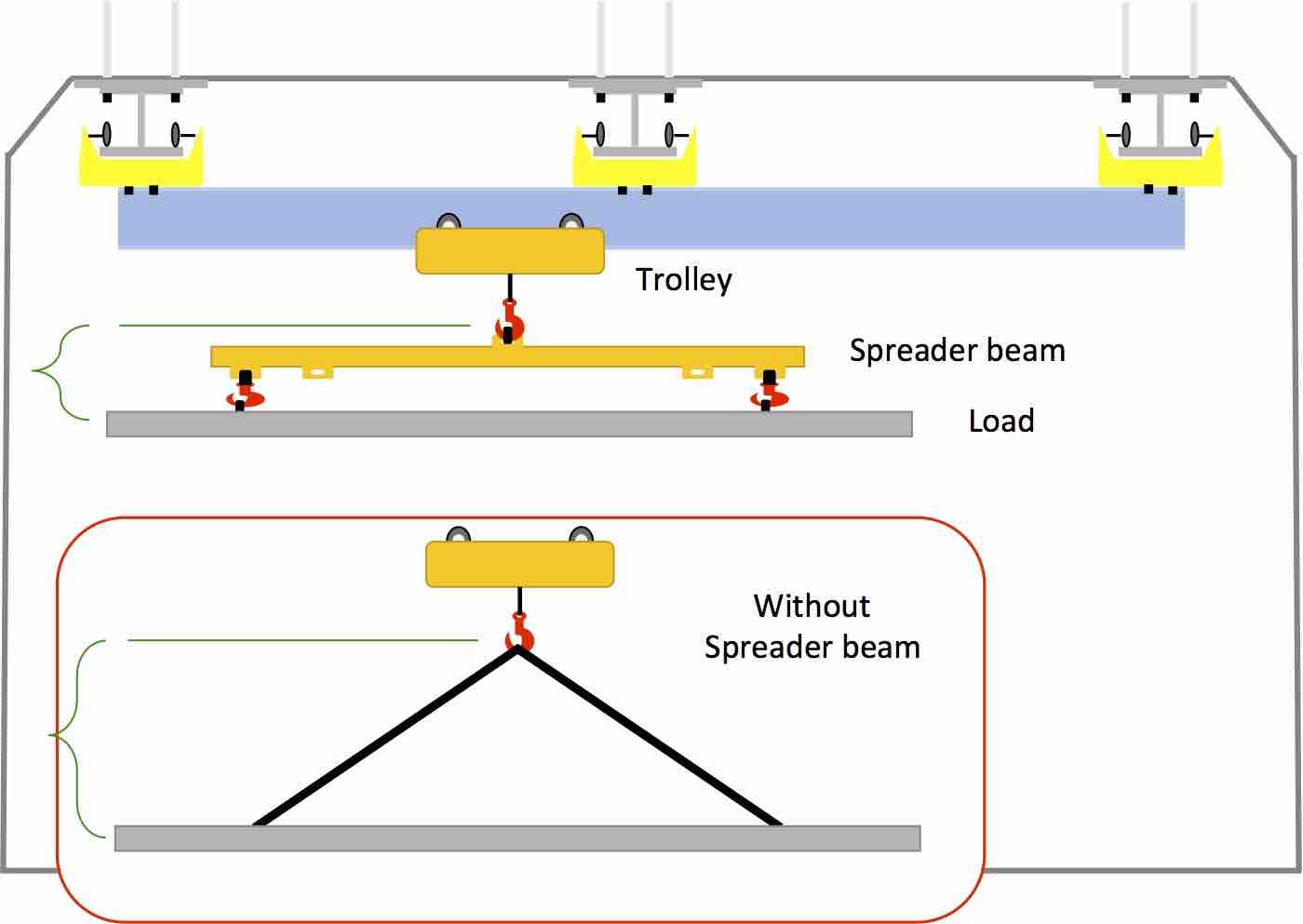}
\caption{{\footnotesize Sketch of a spreader beam, used to minimize vertical height when
rigging long pieces.  This beam is attached to the hoist eye, and to the lifting points of the
load with shackles. The inset shows schematically how vertical height is saved using this device.}
\label{spreader}}
\vspace{0.2in}
\end{figure}

Rigging for the longest -- 10-m -- roof blocks which are closest to the hoist, will determine the minimum clearance required between shielding and hoist, hence the height of the cavern. 
Long pieces are normally lifted from two points, about the 1/4 and 3/4 points, with a single hook
it is necessary to run a cable between these two lifting points and the hoist hook.  
The angle of this cable (referred to as the ``fleet angle'') to support the weight, 
adds considerably to the height needed to lift
the load.  To mitigate this, a spreader beam is used, substantially reducing the distance between
the hook and the load.
As seen in Figure~\ref{spreader}, a spreader beam is an I-beam with 
picking-eyes welded to the outsides of the flanges.  
The top flange is shackled to the hoist hook by a picking-eye in the center.  
Multiple picking-eyes on the bottom flange accommodate different piece lengths.  
The spreader beam is connected to the load using two shackles at either end, 
mid-points can be added as necessary.  Steel and precast concrete can be 
easily equipped for this type of picking.  

The shielding blocks for sides and bottom will be easier to handle.  
The blocks can be made much shorter and more headroom is available.  
The concern then becomes neutrons escaping through gaps where the blocks meet.  
Pieces being carefully sized and shaped will minimize this effect by not providing 
any direct paths for neutrons to take through the shielding.  
The bottom shielding will also need removable blocks to provide 
a passageway to access the ion source.

\subsubsection{Assembly of Target Shielding}

As will be seen in Section 4.1.2, the shielding design for the target area
is largely completed.  
We have a conservative design that meets the requirements for 
rock activation, which we anticipate can be further optimized to reduce its size even more.
Shielding between the target/sleeve and KamLAND steel vessel still requires optimization
to ensure that background levels within the KamLAND detector are not adversely affected, 
but it is anticipated that requirements can be met without substantially increasing the size,
and composition of the already-calculated shielding.

  The target shielding is much more compact than the cyclotron vault
  walls, consequently it contains many fewer pieces.  
  The current design for the target shielding consists of about 110 tons of steel, and 
  about 100 tons of concrete.  
  This would be a total of 21 pieces that are assembled right around the target and sleeve
  assembly.

\clearpage

\chapter{Utilities and Environment}

This chapter will cover the required utilities, electrical power and cooling water, as well as provisions
for supply of ventilation air at suitable temperature and humidity levels.  It will also address 
the control of moisture inside the technical areas.  The radiological aspects of air and water 
leaving the experimental areas will be addressed in the following chapter.

\section{Electrical Systems}

\begin{figure}[t]
\centering
\includegraphics[width=5in]{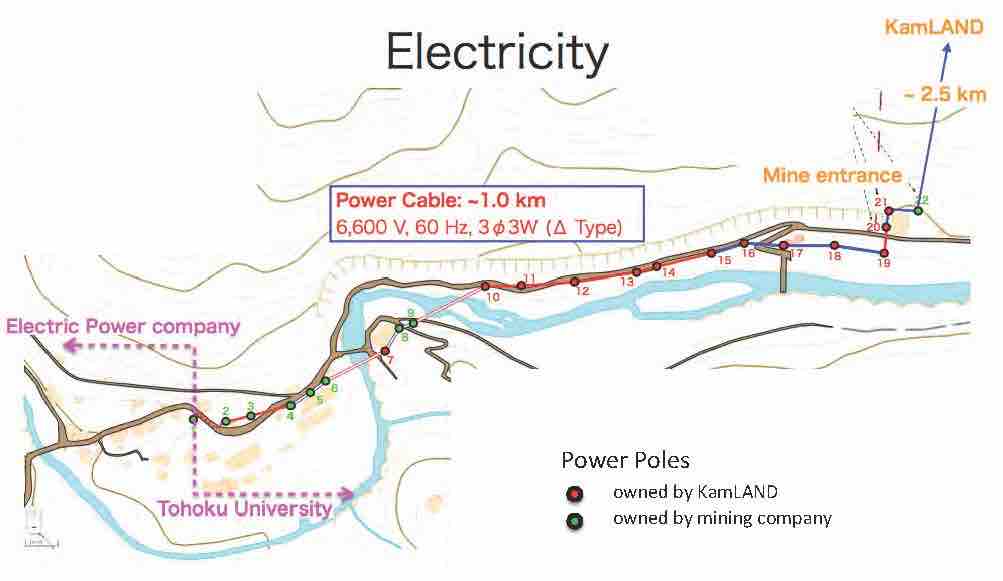}
\caption{{\footnotesize  Routing of power lines is shown by the dotted purple line.
Poles, shown in green and red are owned by the Mining company and KamLAND, respectively.}
\label{external-power}}
\vspace{0.2in}
\end{figure}

Currently, KamLAND has installed power for 1.5 MW.  The location of the substation is 
 shown in Figure~\ref{KamMAP}, along the main access drift.  
The principal load is the purification systems, which are used only on occasion.
Electrical power, at 6.6 kV, is brought from the main power distribution point by 22 poles, 
distributed as shown in 
Figure~\ref{external-power}.

These power poles are owned by the mining company (numbers 1-6, 8, 9, 22) in green numbers) and
by KamLAND (numbers 7, 8 - 21) in red numbers.  Three 60 mm$^{2}$ conductors bring the 3-phase
60 Hz current to the mine.  Extensions of these cables bring power to the KamLAND site.

The requirement for powering the IsoDAR systems is 3.5 MW.

The existing 60 mm$^2$ cables will only transmit about 1.8 MW, therefore two new sets of these same
cables, or slightly larger ones, would be required to power IsoDAR.

To run these additional cables, the following would be required:
\begin{itemize}
\itemsep-0.1em
\item Permission from the mining company to use their poles,
\item Evaluation as to whether the existing poles, both those belonging
to the mining company, and those that belong to KamLAND, are strong enough, and 
can be configured to
support the new cables,
\item As this line runs across the Atotsu river in two places, as well as crossing the public
road, permission is needed from the Hida City office.
\end{itemize}

An additional concern is that the current capacity of the main power line from the 
electric company is not able to supply an additional 3.5 MW. 
It will be necessary to petition them to increase their delivery capability,
which, while not a technical problem, would require a year or two to put into effect.
Note must be taken then, that as soon as the go-ahead to begin construction of
IsoDAR is given, the electric company must be notified of the increased load requirement.

The new power cables must be brought into the KamLAND area, along the same right-of-way
as the present cable, and connected to a new substation capable of supplying 3.5 MW 
to the cyclotron and other IsoDAR systems.

While the detailed layout of transformers, breakers and distribution nets
 for this substation has not been performed,
an estimate for its footprint has been made, based on similar installations.  
As seen above, it is believed the equipment will fit within a rectangular area
that is 15 meters by 7 meters.  
This amount of space has been reserved in the cyclotron area for this substation.



\section{Cooling Water Systems}

\begin{figure}[t]
\centering
\includegraphics[width=5in]{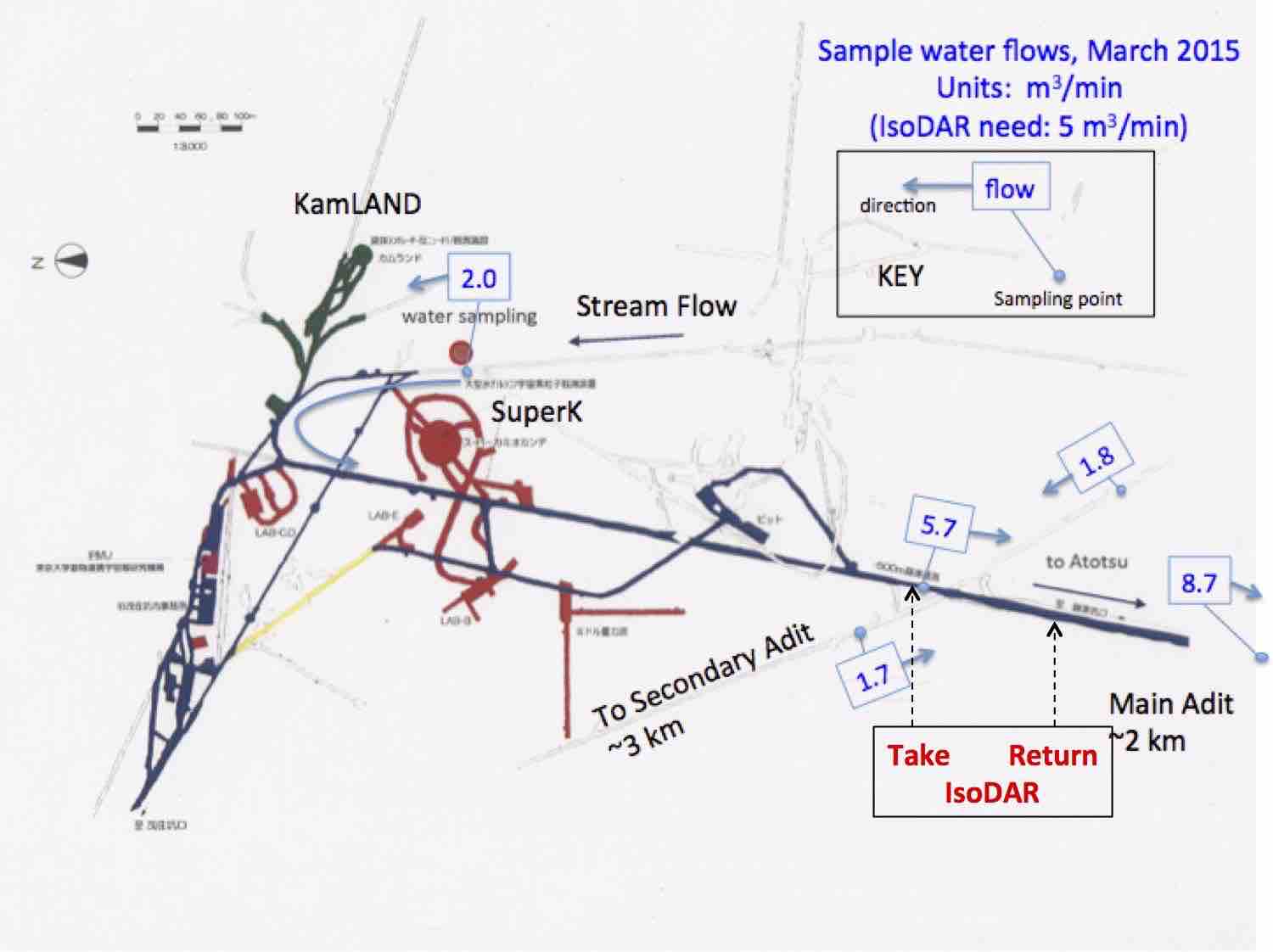}
\caption{{\footnotesize  Route of underground stream that provides cooling for all
existing experiments. Figures in square boxes are measured flow, in m$^3$ per minute. 
Data are from March 2015, which from Figure~\ref{streamStats} is close to a
minimum in historic flow rates.}
\label{stream}}
\vspace{0.2in}
\end{figure}

At present, cooling for all of the experiments is provided by an underground stream
that flows past the SuperK and KamLAND sites, and exits the mine in a trench that runs
alongside the main Atotsu access drift.
Figure~\ref{stream}
shows its route, and the approximate flow rate at given points.

\begin{figure}[t]
\centering
\includegraphics[width=5in]{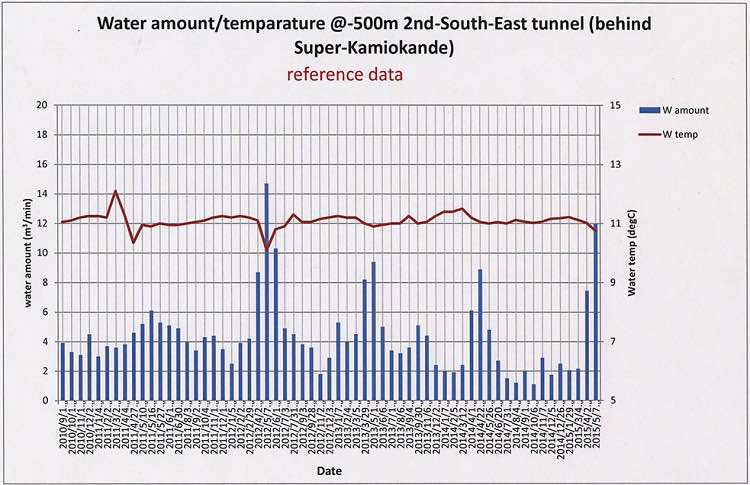}
\caption{{\footnotesize  Measurements of the flow rate, taken at the ``water-sampling'' spot
between KamLAND and SuperK.  One sees large seasonal variations, and in the driest months
there has been worry that the flow will be adequate for existing experiments. It also appears to be
trending downwards.}
\label{streamStats}}
\vspace{0.2in}
\end{figure}

Statistics for the flow rate in this stream are shown in 
Figure~\ref{streamStats}
and indicate a very significant seasonal variation.  April and May represent times of increased 
flow due to snow melting, while the winter months, when ground is frozen, represent low months.

Estimates are that the IsoDAR load requires 5 m$^3$/minute that would result in a temperature
rise of 10\degree C.
While this amount of water is not available in the vicinity of KamLAND, as shown in Figure~\ref{stream}
the flow in the stream is increased as tributaries from other areas feed into the main river,
and at the indicated points there would probably be adequate flow for IsoDAR uptake and return.

It should be noted that SuperK in particular relies on this stream to maintain temperature stability
in its systems, so it is very sensitive to any perturbations in the water temperature.  
For this reason, also, any proposed use of this stream for cooling IsoDAR would require pumping 
water from a point well downstream of the SuperK intake.  Water pipes (25 cm diameter) of the order of 1 km long will be needed to bring this water to the IsoDAR site.

However, even if the 5 m$^3$/min flow is available, the temperature rise of 10\degree C may be 
too high to allow direct discharge of the water into the Atotsu river.  
This will need to be checked, and if true then an auxiliary cooling tower will be
required to reduce the temperature of the water prior to release into the river.

There has been concern with the reliability of this stream, especially since KAGRA excavation
seems to have affected the flow paths of water in the mine.

A preferred option, then, would be to use the cooling tower, that will be most probably
required in any event, and run water lines to IsoDAR so providing a closed-loop system that is 
totally independent of the stream.  An extra 3 km of pipes will probably be needed.

\begin{figure}[t]
\centering
\includegraphics[width=5in]{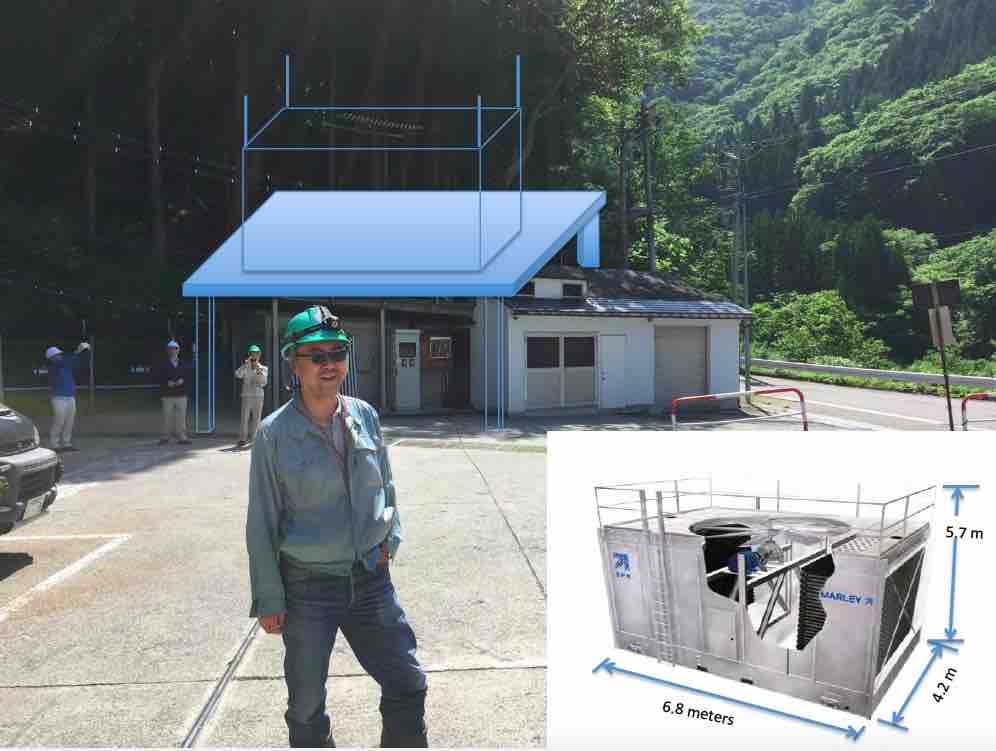}
\caption{{\footnotesize  Possible location of the 3.5 MW cooling tower, 
on a platform above the air-supply shacks providing dehumidied air 
to KamLAND (left side) and SuperK (right side).  
An alternate site for the cooling tower would be across the road, 
where more flat area is available.  
The cooling tower measures 4.2 x 6.8 x 5.7 meters.}
\label{coolTwr}}
\vspace{0.2in}
\end{figure}

Figure~\ref{coolTwr} 
shows a possible location for this cooling tower, on a platform on top 
of the air supply buildings at the entrance portal to the mine.  An
alternate site would be on flat terrain on the other side of the road,
as there is substantially more space there.

The cooling tower would need to be sized for 1000 tons of refrigeration capability, for the 3.5 MW of heat dissipation required.  An example is Marley Model NC8412V-1, with a footprint of 4.2 by 6.8 meters and a height of 5.7 meters.  

Approximately 4 km of 25 cm steel water pipe must be installed,
 and water pumps at either end
of the line will be needed to drive the water through the circuit.


It is possible, if this system is available, its reliability may be attractive to other experiments, 
with a possibility of some cost sharing.

\section{Input Air Quality and Humidity Control}

\begin{figure}[t]
\centering
\includegraphics[width=6in]{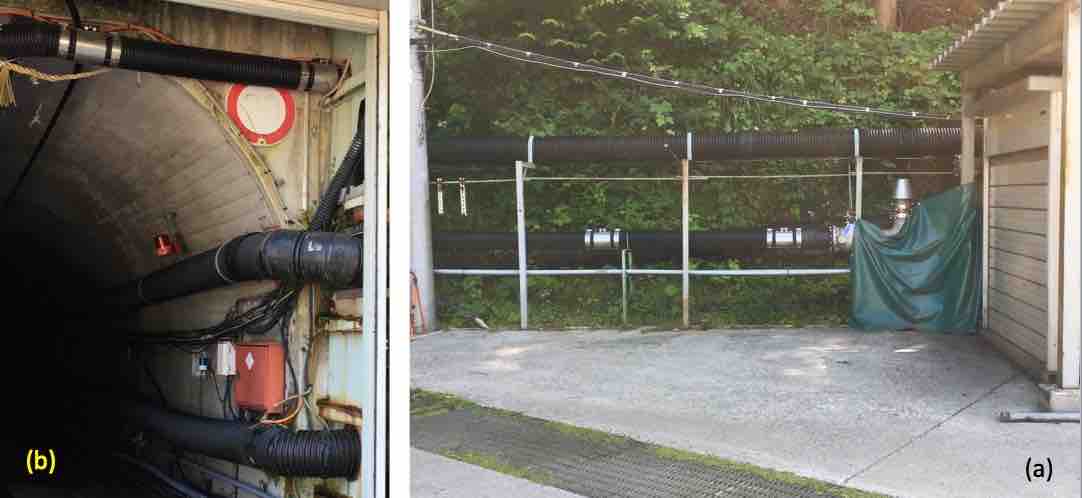}
\caption{{\footnotesize  Air lines for KamLAND (lower) and SuperK (upper), (a) near pump
buildings, and (b) entering mine access road.}
\label{airLine}}
\vspace{0.2in}
\end{figure}

KamLAND provides dry, de-humidified air to its critical areas, especially the top
of the detector dome.  
This air flow is generated in a small building, shown in
Figure~\ref{coolTwr}
underneath the sketch for the location of the cooling tower.  
The larger building on the right side provides fresh air to SuperK.
The dehumidified air is transported to the KamLAND area via a solid pipe,
about 30 cm diameter, the bottom black pipe in 
Figure~\ref{airLine}.
is the KamLAND line, the upper one is SuperK's.

This system is not able to provide air to both KamLAND and IsoDAR, but a duplicate system 
would meet the needs of IsoDAR.  This would require a separate blower and dehumidifier installation.
A second flexible trunk line could be installed, identical to the current KamLAND run, 
or if space is tight along the main access roadway for a second air line,
the KamLAND line could be replaced by one of larger diameter 
that would carry twice the airflow.


\section{Moisture and Environmental Water Management}

\begin{figure}[t]
\centering
\includegraphics[width=5in]{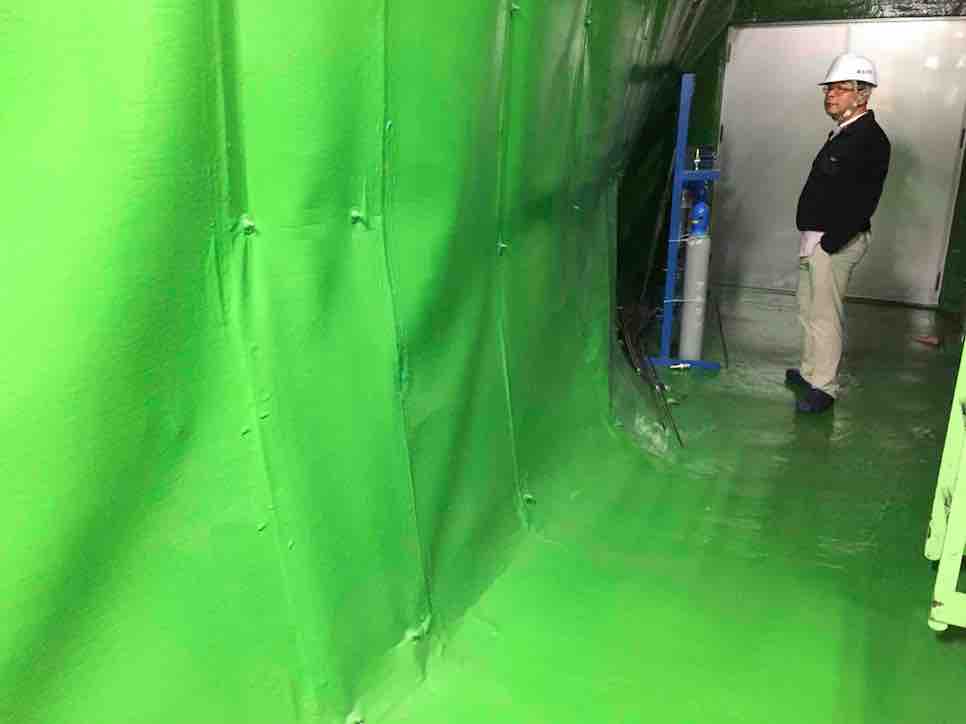}
\caption{{\footnotesize  Polyethylene sheet, approx 2 mm thick, lines the rock walls,
floor and ceiling, 
in areas where cleanliness and environment control is important.  The KamLAND xenon storage and liquid scintillator purification areas are lined with this material.}
\label{polyLiner}}
\vspace{0.2in}
\end{figure}

The environment of the target and cyclotron areas must be controlled to ensure cleanliness, 
protection from moisture, and isolation for radiological considerations.
Figure~\ref{polyLiner}
shows the technique employed by KamLAND for this isolation. 
A polyethylene sheet, about 2 mm thick, is attached to the walls and ceiling,
and is glued to the floor, providing a completely sealed room.  

This system
should be employed for the IsoDAR target and cyclotron areas, as it provides
excellent control over the internal environment in these areas.
Not only does it ensure that water will not penetrate to the critical components, 
and air temperature and humidity can be controlled,
but it also provides a radiological barrier to prevent contamination of external areas.
This will be discussed at length in the next chapter.

One potential complication is that it is necessary to maintain the areas inside
these barriers at a slight negative pressure, in order to contain any contamination
products, mainly in air, that might be made by stray neutrons.  This will require
special engineering studies to ensure proper adhesion of the membrane to the wall
so the membrane does not collapse under this negative pressure.

\clearpage

\chapter{Radiation Protection\label{radpro}}

This chapter will address aspects of radiological protection related to the IsoDAR
experiment installed at the KamLAND
site.  This is a particularly challenging subject, as the entire experimental program in the 
Kamioka mine is based on obtaining the lowest-possible radiation background. 
At a depth of 2500 meters-water-equivalent, the attenuation of cosmic ray muons
is about a factor of 10$^6$, providing suitably low rates for the high-energy muons
that do reach the large detectors.  
So introducing neutrons and artificially-produced radioactivity is a significant issue.

Radon is always an issue in underground environments;
 the uranium and thorium content of the rocks have been well
characterized, at a few parts-per-million of each, sufficient to produce significant
amounts of radon unless protective measures are taken.
Where necessary, barriers are employed to mitigate radon penetration to sensitive areas,
mainly via clean rooms with controlled air flow (and slight positive pressure) and exceedingly high filtration.

For the high-power IsoDAR cyclotron, capable of producing 600 kW of 60 MeV
protons, the primary risk comes from
neutrons produced in nuclear reactions of the protons.  Neutrons are 
difficult to contain, and are readily captured by materials such as the rock of the
cavern walls.  Other materials can also absorb neutrons, and an 
analysis must be performed to establish possible activation of ground water
that might flow in close proximity to the radiological sources, or even air in the
caverns where neutron fluxes are high.

This chapter will first address the shielding necessary to keep neutrons from reaching
rock in the caverns, and to maintain activation levels below those required
by Japanese regulations.  We will also explore strategies for addressing
possible air and ground-water activation.

As the target will be in close proximity to the KamLAND detector, shielding must also satisfy
the requirement of not unacceptably increasing the background level of neutrons and gamma 
radiation in the sensitive detector volume.  As this endpoint for shielding is different from that 
of pure rock activation, new calculations are needed once background studies are performed
to determine the neutron attenuation levels needed.  It is likely that thicker shielding may be 
needed in the direction of the detector.  This work will be performed in the coming months.

The chapter will end with a brief discussion of personnel radiological protection, relating
to the KamLAND and IsoDAR staff that require access to the site, and 
instrumentation to ensure adequate personnel protection.

\section{Shielding and Rock Activation}

\subsection{Requirements}

The primary shielding requirements for IsoDAR are different from a normal
radiation-producing accelerator, in that the critical element
that must be protected is the surrounding rock, rather than personnel 
present in the vicinity during accelerator operation.

Of course, personnel must be protected, however there is little need for
personnel to be close to the cyclotron or target when beam is being produced,
 and their protection
can be handled by administrative measures, namely exclusion zones.

For personnel, the important consideration is the instantaneous total radiation dose
(neutrons and gammas)
at the surface of the shielding, while for the rock it is the integral
dose (of neutrons only) over the full time period of the experiment.  The energies of 
the gamma rays are too low to cause nuclear transmutations, so these are
of no concern in the shielding calculations.  However, neutrons are readily
absorbed, creating radioactive isotopes.

Japanese regulations regarding activation of environmental material follow the requirements
specified in the IAEA 
(International Atomic Energy Agency) Safety Standard 
Series, No. RS-G-1.7, ``Application of the Concepts of Exclusion, Exemption and Clearance.''
Table 2 of this document lists threshold values for artificially produced radioactivities;
essentially all the isotopes relevant to neutron activation of rocks are limited to values
less than 0.1 Bq/gm.  (These isotopes will be listed below.)

Discussions with radiological protection officers at RIKEN have helped evaluate how
these limits affect the IsoDAR experiment.  
\begin{itemize}
\itemsep-0.1em
\item Concentrations above the established limits must be contained in a 
Controlled Area for Radiation Protection, and
\item During operation of IsoDAR there will be well-defined
Controlled Areas where personnel access must be carefully monitored and regulated
 because of high
radiation fields from the accelerator, transport line and target, 
\item When the experiment is finished, and all the equipment has been removed,
there must be no remaining Controlled Areas.
\end{itemize}

Thus, at the point when the experiment has been declared to be complete, 
the maximum level of (the sum of) all artificially-produced isotopes anywhere
in the caverns where the equipment was located must be less than 
the 0.1 Bq/gm limit.
Following the end of beam-on-target, a substantial time period should
elapse to allow the short-lived activity in the experimental components to decay.  This time
delay will
reduce the exposure of personnel working in the area, when dismantling of the
experiment begins.
It should be assumed that essentially all of the experimental apparatus and shielding 
material will be radioactive, so dismantling and disposal of this equipment should
be done slowly and deliberately.  The removal of the last piece
designates the ``end of the
experiment.''  At this point, the assay should be performed to establish
that activity remaining in the walls is low enough that no controlled areas
remain.

\begin{figure}[t]
\centering
\includegraphics[width=5in]{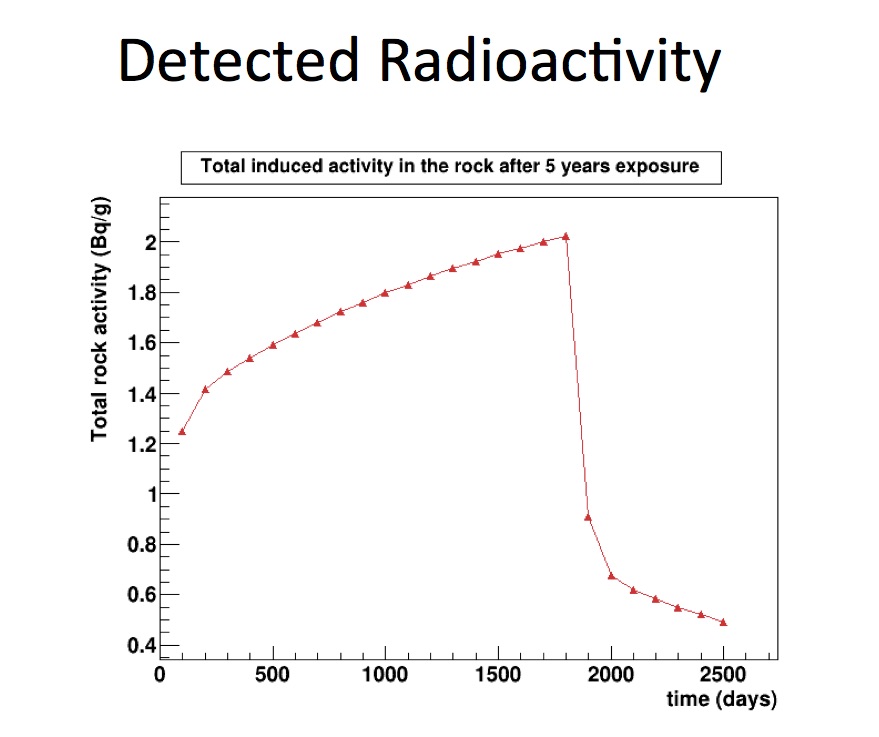}
\caption{{\footnotesize  The curve shows the buildup of activity in the surrounding 
rocks for the 5-year planned beam-on-target.  Once beam is shut off, rapid decay
of the many short-lived isotopes occurs, leaving only long-lived activities.  The 
``end of the experiment'' is declared at approximately day 2500, by which time the
activity level has dropped very significantly.}
\label{decayCurve}}
\vspace{0.2in}
\end{figure}

Figure~\ref{decayCurve}
shows how the activity in the rock decays after the source of neutrons is turned off.  
The steep drop right at the end of beam represents all the short-lived isotopes 
that quickly decay away, leaving only the long-lived ones.  
Table~\ref{isotopes}
lists the relevant isotopes and their half-lives.  All but one are produced by 
capture of thermalized neutrons, in trace amounts (parts-per-million) of parent source material
in the rock.  The exception is $^{22}$Na, produced by high-energy neutrons (the threshold
energy is 11 MeV), but about 6\% of the rock material is sodium, so a very few
remaining high-energy neutrons can create a large activation problem.

\begin{table}[!t]
\caption{Long-lived isotopes produced in rock. (*TNC = thermal neutron capture)\label{isotopes}}
\centering
\renewcommand{\arraystretch}{1.25}
\begin{tabular}{lllll}
\hline
Radionuclide & Half life & Parent & Cross section & Produced by\\
	&	& (Abundance) & (barns)	&  \\
	\hline
$^{60}$Co & 5.3 years & $^{59}$Co (30 ppm) & 37 & TNC* \\
$^{152}$Eu & 13.5 years 	& $^{151}$Eu (3 ppm) & 4600	& TNC \\
$^{154}$Eu & 8.6 years & $^{153}$Eu (3 ppm) & 310 & TNC \\
$^{134}$Cs & 2.1 years & $^{133}$Cs (30 ppm) & 29 & TNC \\
$^{46}$Sc & 84 days &  $^{45}$Sc (5 ppm) & 27 & TNC \\
$^{22}$Na & 2.6 years & $^{23}$Na (6\%) & 0.1 & (n,2n) $<$11 MeV\\
\hline
\end{tabular}	
\end{table}

The abundances of parent elements (except sodium) were established by irradiation and activation
analysis at the MIT reactor of 
rock samples taken from the actual KamLAND site.  Sodium content cannot be established by
thermal neutron activation, however geological evaluation of the rocks from earlier studies
has established the 6\% value for its concentration.

The requirement, then, is that when the radioactivity assay is performed, the sum
of all the activities listed in Table~\ref{isotopes} must be less than 0.1 Bq/gm.
One should note that the assay must be performed with a high-quality instrument,
capable of identifying gammas from each of the isotopes,
as the background radiation from natural uranium and thorium in the rock is almost
50 times higher.


\subsection{Shielding Calculations:  Target}

To perform shielding calculations, it is necessary first to establish a ``source term.''  That is,
how many neutrons are produced at the particular location, that must be prevented
from reaching the rock.

In the case of the target, the source term is easy to determine.  The experiment is based
on production of the maximum possible number of neutrons in the beryllium
target.  This neutron flux establishes the $^8$Li production, and thus the
number of antineutrinos.  
Over the 5 years of the experiment, it is estimated that a total of 8.5 x 10$^{23}$
neutrons will be produced.
For a sphere of 2.5 meters' radius, this translates into about 10$^{18}$ neutrons
passing through each square centimeter on the spherical surface.
From this, and the abundances and cross sections of parent materials in
Table~\ref{isotopes},
one can evaluate the shielding necessary to stay below the 0.1 Bq/gm requirement.

\begin{figure}[t]
\centering
\includegraphics[width=5in]{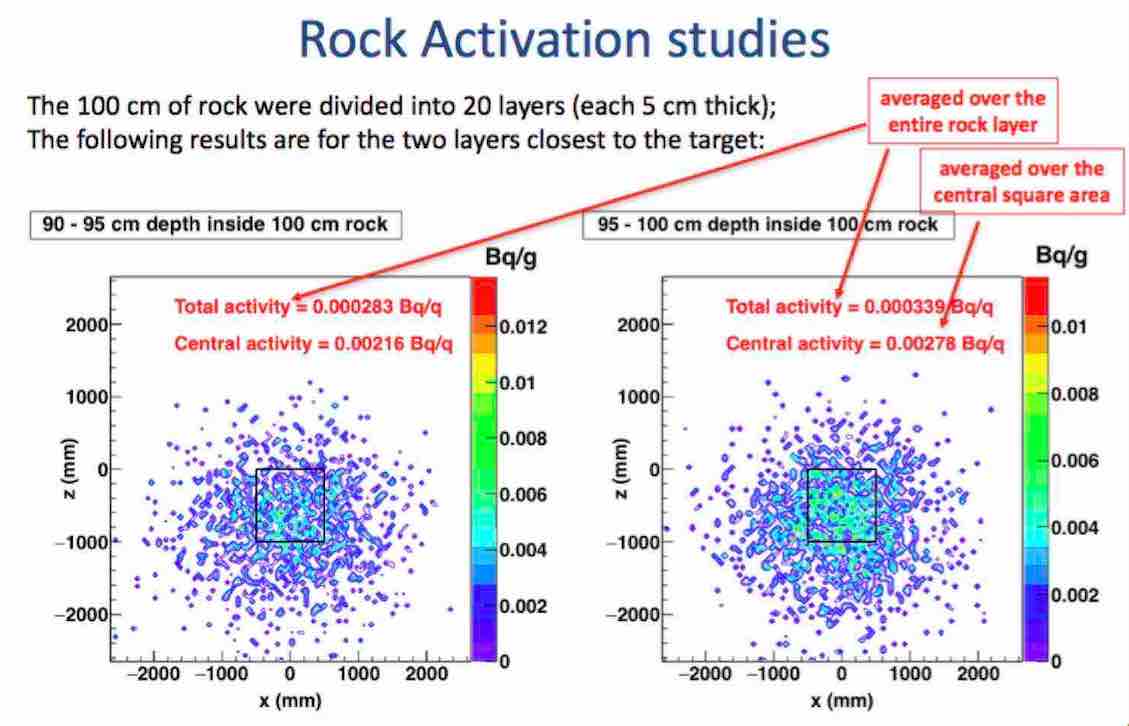}
\caption{{\footnotesize Total induced activity in the rock layers closest to the surface
of the shielding.  This is an (x,y) plot, the beam is coming from the top of the figure 
heading downwards.  The highest concentration of activity is at the very center, which
is the 90\degree direction from the target, so is the thinnest part of the shielding.
Other coordinates have thicker shielding, depending on the cosine of the angle
from the target.  It is clear to see that this configuration of shielding, namely 1 meter
of steel and 1 of boron-loaded concrete, easily meets the 0.1 Bq/gm activation requirement.}
\label{activation}}
\vspace{0.2in}
\end{figure}

GEANT4 is the simulation code we use to evaluate shielding requirements.
This code, with appropriate well-calibrated libraries of cross sections, has been 
extensively benchmarked, and is used widely around the world.  
The results obtained are also being cross-checked by the RIKEN group, using the PHITS code, 
the primary shielding code
used in Japan.  
We do not have cross-comparisons yet, however this work is ongoing and we should have
such comparisons in a few months at most.

The most recent results with our GEANT4 code are shown in 
Figure~\ref{activation},
for a shielding configuration consisting of 1 meter of steel and 1 meter of concrete with a special 
aggregate of boron carbide, particularly good for absorbing slow neutrons.  
For these calculations, the neutron spectra emerging from the top of the shielding was determined,
and these neutrons then were transported through one meter of rock, divided into 10 cm slices.
The highest activity level is registered in the layer closest to the target, labeled 95-100 cm.
The  densest area of activity represents the thinnest part of the shield, namely the direct 90\degree
 line to the target.  As one goes away from the center the shielding thickness increases, as the cosine
of the angle.  
The estimation of the activity level, for all the isotopes tracked (those in Table~\ref{isotopes}), is 
clearly substantially below the 0.1 Bq/gm requirement, in fact in the worst case, the level is a factor of 40 below this.

Meeting the stringent activation requirement is
certainly possible.   For the PDR, we might be able to optimize the shielding design
to reduce the amount of material, as well as decrease the amount of rock that needs
to be excavated to fit the shielding.    At that time we will have completed
the PHITS intercomparison, lending additional confidence in the numbers generated.

\subsection{Shielding:  Cyclotron}

The ``source term'' for neutrons from the cyclotron is more difficult to evaluate.
Neutrons all come from beam loss, which will be minimized as much as possible.
However, such losses cannot be brought to zero.

A practical and commonly used guideline to keeps the beam losses inside the cyclotron vault
and transport lines to less than 200 watts.  This number, developed at PSI,
results in activation levels of components that do not preclude hands-on maintenance.
For calibration, 1 watt of beam loss will come from 1 microamp of beam at 1 MeV.
Most of the beam loss we will experience will be at the full energy of the cyclotron,
60 MeV, so 200 watts will be about 3 microamps or 1.8 x 10$^{13}$ protons/second.
As the full beam is 10 milliamps,
this represents a loss of 1 part in 3000 of the primary beam.  This level of beam loss
is achieved at the 590 MeV cyclotron at PSI.   We are
working closely with the PSI group, to benefit from their experience.

Losses can be characterized as ``controlled'' and ``uncontrolled.''
``Uncontrolled'' losses arise from beam interacting with residual gas in the accelerator
and transport lines, from beam halo, namely trajectories that take the
beam particles outside of the normal stable orbits, or any other mechanism that causes particles to
strike surfaces inside the vacuum tanks.
The radiation resulting from this type of loss is distributed fairly uniformly around
the entire inner surfaces of the vacuum enclosures, so is difficult to shield.

``Controlled'' losses, on the other hand, are expected, and are channelled into 
well-shielded areas.  In our case, it is inevitable that some beam will be lost
on the extraction septum, possibly as much as a few kilowatts.  This loss occurs
because it is impossible to have completely clean separation between turns (orbits) 
in the cyclotron, so a small number of particles can be found in the valley 
between the last turn to circulate, and the turn that enters the extraction channel.
This channel is defined by a thin septum, a few micrometers thick of graphite, usually.
This septum is one electrode of a high-voltage plate system that provides a kick
to the beam so it exits the cyclotron.

\begin{figure}[t]
\centering
\includegraphics[width=5in]{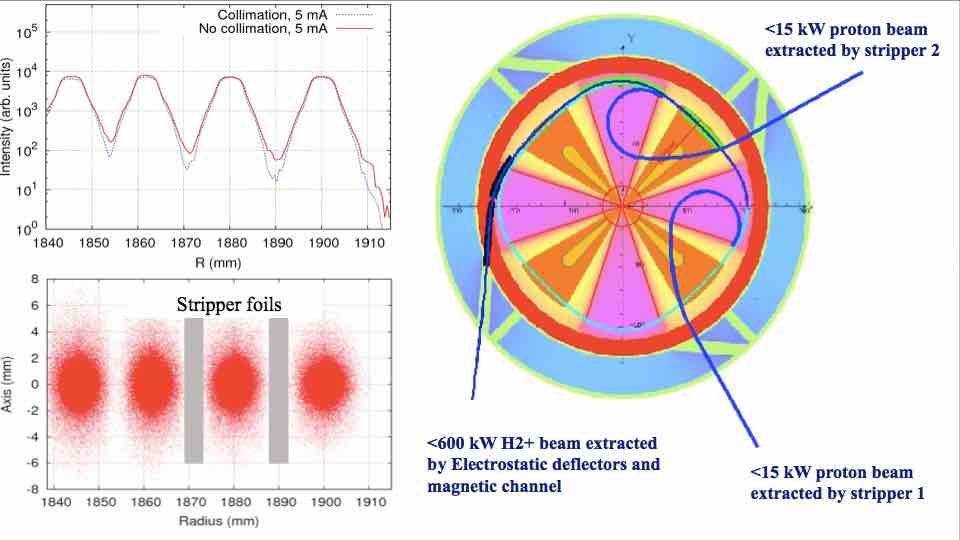}
\caption{{\footnotesize Orbits of protons emerging from thin stripper foils designed to clean
up beam halo and keep \htp ions from striking the extraction septum.  As is seen, these protons
can be made to cleanly leave the cyclotron.  Key:  pink = hill field $(\langle  B{_{Hill}}\rangle  \approx  2{\rm T})$, yellow = valley field
 $(\langle B{_{Valley}} \rangle   \approx$ 0.02T), orange = RF Dee. The valley field includes effects from the fringe fields of the hill poles. }
\label{StripperOrbit}}
\vspace{0.2in}
\end{figure}

As the beam circulating in the cyclotron is H$_2^+$, it is possible to protect this
very thin septum with a narrow stripper foil.  The foil intercepts
all particles that would strike the septum, and breaks up the fragile H$_2^+$
molecule into two protons.  The magnetic field between the stripper and the
septum must be sufficient so that the protons are bent enough to clear the inside
edge of the septum.  As seen in 
Figure~\ref{StripperOrbit}
stripper foils placed at the entrance to a hill bend protons around, and the weaker field in the valleys 
brings them out almost perpendicularly to the H$_2^+$ orbit.  The figure shows two such channels,
one on the turn prior to extraction, the second on the extraction turn itself.  These protons
pass into well-shielded beam dumps, designed to
absorb the protons and also as many as possible of the neutrons produced by
the stopping protons.  It is anticipated that this "waste" beam could even be utilized
for production of radioisotopes, though this is outside the scope of
this CDR.

\begin{figure}[t]
\centering
\includegraphics[width=5in]{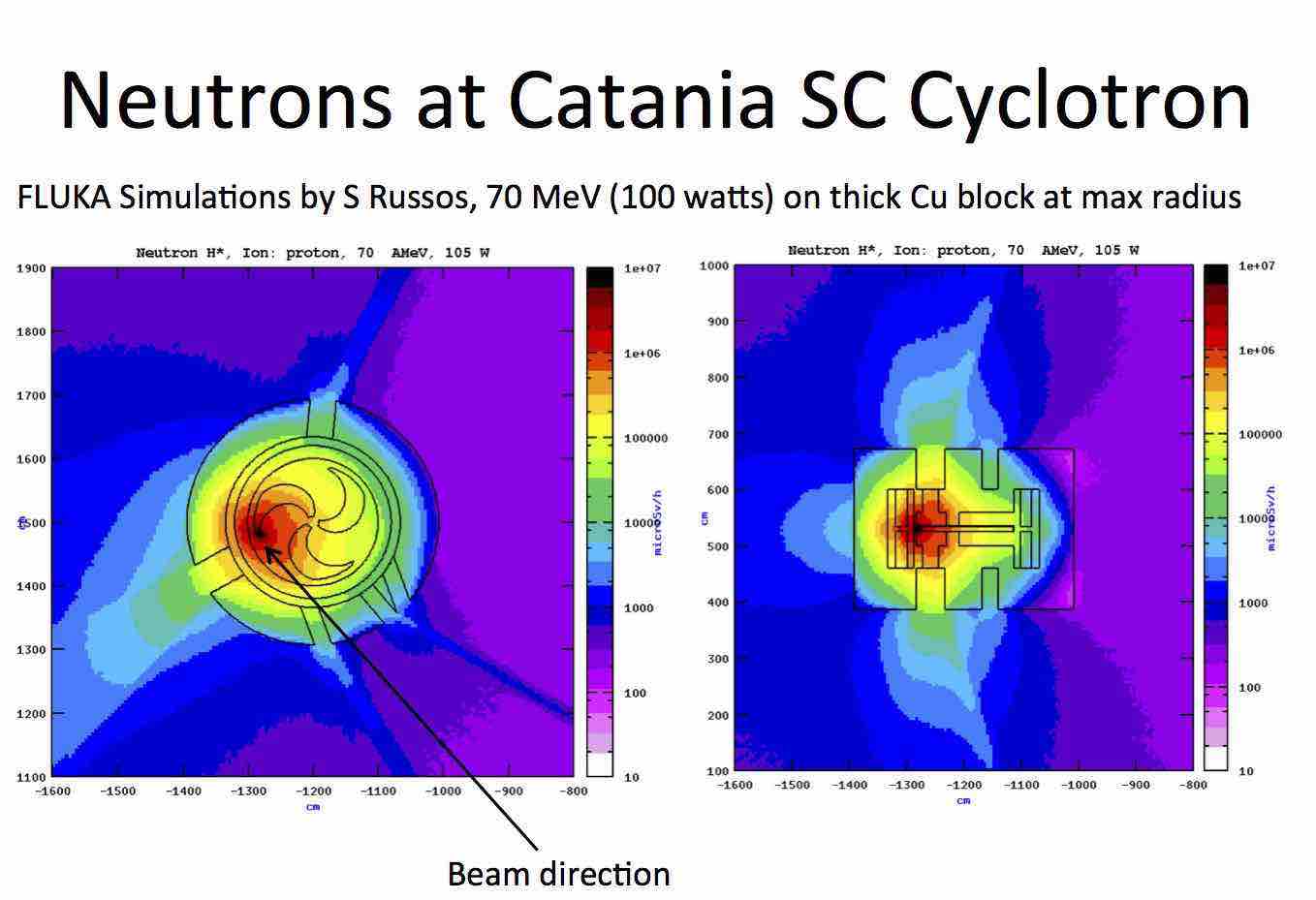}
\caption{{\footnotesize FLUKA simulation of neutron flux around the Catania 70 MeV 
superconducting cyclotron.   Left is the top view, right is the side view.  The simulation
assumed 100 watts of 70 MeV protons strike a copper block, at about the location of
the extraction septum.  One can see that the bulk of the steel of the cyclotron contains
most of the neutrons, however areas where the steel is thin: extraction port (about 6 o'clock in 
left figure) and the RF dee stems (above and below the plane in right figure) do show significant
neutron dose levels. }
\label{cyclotron-neutrons}}
\vspace{0.2in}
\end{figure}

Characterizing and quantifying uncontrolled losses is a difficult problem.
Simulations can be performed,
Figure~\ref{cyclotron-neutrons}
shows one such calculation, done with FLUKA for the geometry of the Catania
superconducting 70 MeV cyclotron.  This figure demonstrates the complexity of
the problem.  The very large amount of steel in these machines provides a 
significant amount of self-shielding, however there are thinner parts, particularly 
for RF dee structures, that provide paths for neutrons to escape.
One can perform similar calculations for the designed geometry
of our cyclotron, and doing these will give us good data for optimizing shielding
design.  However, these calculations have not yet been initiated.

Therefore, as a worst-case scenario, we can take the case all the
beam is deposited onto a neutron-producing target for five years.  We also have established
that for this case a shield of 1 meter of steel and 1 meter of concrete provides
adequate protection.
So as a starting point, if we specify that the entire cyclotron is surrounded by such
a shield, we know that rock activation will not be an issue.

This make the cyclotron cavern larger than necessary, and will also 
substantially overestimate the inventory of shielding material.   But starting here will
give us an upper bound on the configuration of the experiment.  Sizes and costs will only
go down when we refine the calculations.

\subsection{Shielding:  Beam Transport Line (MEBT)}

Beam is extracted from the cyclotron as H$_2^+$, but we choose to strip it as soon as
possible into protons.  This will minimize uncontrolled losses in the MEBT, which would mostly
come from stripping of the fragile H$_2^+$ molecules by residual gas in the transport line.

The stripper is located a few meters from the extraction port, and is close enough that it
can be included in the cyclotron shielding vault.  The stripper is followed by an 
analysis magnet and beam dump for unstripped beam, and halo-cleansing collimators.  These
are also all located inside the cyclotron vault, so we should be comfortable with any neutron
production being adequately shielded.

Transport of the beam from the cyclotron vault to the target area, and its very adequate shielding,
should be clean, with little or no beam losses.  However, space is available for some local
shielding along the beam line should it prove necessary.  

Calculations need to be done of the effect of beam losses in this region, either from mis-steering
during tuning, or momentary power-supply failures.  However,
as indicated, it is not expected that providing shielding in the event it proves advisable, will
be a major issue nor will it contribute significantly to the costs of deployment of the experiment.

\section{Control of Activated Materials}

As in any accelerator facility that produces ion beams with energies above the Coulomb barrier, there is always a chance
of  activating any pieces that may be exposed to the beam or neutrons produced by the beam.  
The measures for handling such materials are well-established, and will be adopted
for our experiment.
One item of great importance will be monitoring for leaks in the primary cooling circuit for the target.
Because of the intense neutron flux that this water is exposed to, it will contain
activated material.  Primary among these will be tritium.  However, other isotopes, such
as $^{15}$O, $^7$Be, and several others will undoubtedly be present.  

Design of 
water circuits to operate in this environment are well understood, they are called RAW 
(RadioActive Water)
systems.  Good experience exists at Fermilab with such systems, and no doubt experienced engineers
at J-PARC and KEK have designed and operated such systems as well.  
Procedures for the design and maintenance of our RAW cooling circuit will be adapted from the other
laboratories with experience with such systems, these will render
highly unlikely, or in the worst case, contain any leaks that might occur.

Should a leak develop during normal operations, system instrumentation would detect a large leak
through water pressure drops.  Small leaks would probably appear through air contamination of
evaporated water containing trace levels of tritium or other products.  
The systems put in place to monitor air exiting the high-radiation areas would pick up this signal.
This will be discussed in the next section.
But in any event, the impermeable polyethylene lining of the caverns would contain any leaked
water, so there would be no penetration into the rock surfaces.  Cleanup would occur
using standard decontamination procedures.

Other activated pieces will come from maintenance activities on accelerator components, 
in particular areas inside the cyclotron that will need to be replaced.  Best examples are: the
extraction septum, the stripper foil assemblies, both internal and external.  Other pieces, such as 
the spiral inflector, internal cyclotron probes, and beam line instrumentation are among those 
that will undoubtedly require service.
(Beam lost in the central region of the cyclotron will be at too low an energy to cause activation, 
but the spiral inflector and other central region components will be directly exposed to neutrons
produced at the outer extremities of the cyclotron, so are likely to be activated.)

Handling of these activated pieces will also follow standard procedures of accelerator
laboratories:  localized shielding if levels are too high, limiting time for personnel to work
in these areas, and well-shielded lockers or containers to store and transport the
activated pieces for proper disposal.

Low levels of radioactivity are likely to be seen in vacuum pump oil.
When servicing
these pumps, special monitoring and handling procedures will be followed.

Personnel working in all of these areas will follow the usual procedures of clean suits, face masks,
protective eyewear, helmets, gloves, hair nets, and will always check out through
radiation detectors suitable for the isotopes likely to be encountered, including tritium.

\section{Exhaust Air Control -- Radiological Aspects}

A small chance exists that neutrons escaping from the shielding may undergo nuclear
reactions with surrounding air in the caverns.  
In addition, airborne contaminants might exist: from discharge of the vacuum pumps
that might collect material from the inside of the vacuum chamber of the cyclotron and
beam lines; or from evaporation of accidental leaks or discharges from the primary 
coolant circuit.  Tritium would be the most likely volatile product of such an event.

The primary mitigation measure will be to maintain the atmosphere in the caverns at
a slightly negative pressure.  This will ensure that air is always flowing into the enclosure, 
and the exhaust, channelled through blowers maintaining the negative pressure, 
can be properly monitored for radioactivity prior to exhaust through a
high air stack outside the mine.

Isotopes normally found in high-energy, high-intensity accelerator environments 
are $^{11}$C, $^{13}$N, $^{15}$O and $^7$Be, the first three produced by (n,2n) reactions,
 the latter from spallation reactions of high-energy neutrons, all on air constituents.  
 These reactions all require high-energy neutrons,  which we make great effort to
 minimize through good shielding.  
 The first three all have short halflives, longest is $^{11}$C (20 minutes), so presence of these
 can be mitigated by storing air a short time before releasing it.
 $^7$Be is not volatile, it can be easily filtered.
 Contaminants from vacuum pump exhausts are also not volatile, and can also
 be filtered.
 Tritium is a more severe problem, however the tolerance levels for release are 
 substantially higher than for other isotopes, and detection of any amount of
 tritium will immediately signal a water leak in the primary coolant circuit which
 will cause immediate shutdown of the system for repair.  This shutdown should
 occur prior to reaching the allowed limits for tritium release.
 
 The strategy followed in other high-power accelerator centers that have similar beam
 characteristics to ours (Catania, Legnaro, PSI), is to maintain their accelerator
 vaults at a negative pressure, monitor the air, run it through several stages of
 filtration, following which it is immediately released, assuming no activity is detected.
  We could follow the same procedures, with good filtration, but then as an added
 measure, conduct the air through a flexible pipe to the surface for discharge in
 an area that is suitably remote and removed from human occupancy.  The long distance
 of travel prior to release will provide further time for any short-lived products to decay.
 
 One point that should be considered:  the caverns with IsoDAR equipment
 will all be lined with the polyethylene sheet described in Section 3.4 above.
 It is important, to maintain the slight negative pressure, that the enclosure be as air-tight
 as possible, and that air entering is carefully metered.  
 This will mean that a pressure differential will exist across the polyethylene membrane,
 with a net force inwards.  
 The stress on this membrane should be calculated, and the application of
 the membrane done in such a way as to not compromise its integrity.

\section{Radiological Aspects of Ground Water Management}

The underground areas around the KamLAND experiment are subject to passage
of water; the amount depending on the season, amount of snowmelt, and other
factors relating to the percolation of water through the rock of the mine.
It is inevitable that water will penetrate into the caverns containing the high-radiation
parts of the IsoDAR experiment, and so water management is essential.

The only water of concern is that which actually enters these
caverns.  By the nature of the shielding requirements, the neutron level inside the
rock walls of these caverns is extremely low, so the probability of activating any
water inside the rock more than a cm or two from the surface, is negligible.
Also the polyethylene sheet prevents water from reaching the experimental
equipment, so the only water we must be concerned with percolates down the walls between
the rock and the polyethylene.  

The mitigation strategy is to collect all this water, and monitor it for contamination prior to releasing it.
To accomplish this, the caverns will be carefully surveyed to determine the entry points of
water, and plot its course through each cavern.  Also, exit points should be mapped, 
and sealed to prevent uncontrolled escape of this water.
As excavations will occur in all the
affected caverns, special channels can be constructed to catch this water and collect it to
a sump where it can be moved to a containment vessel for analysis.

\section{Personnel Access Control}

\subsection{Procedures}

During the course of the IsoDAR experiment, from the beginning of commissioning
of the cyclotron until the end of the experiment, 
the areas around the cyclotron, beam transport and target must be declared, 
at a minimum, as ``Controlled Areas for Radiation Protection,''
and access to these areas must be strictly monitored 
and controlled.    

During actual beam operation, all personnel must be excluded from these areas,
as high radiation levels are likely to be present.  
This radiation can come from x-rays from the high-voltage and cyclotron RF systems, 
as well as from neutrons and gamma rays arising from beam loss in the accelerator
and transport lines.  

When beam is off, however, these radiation levels for the most part disappear.  The
only radioactivity then comes from the decay of components activated as a result of
loss of high-energy protons. 
However, these levels may be high initially, but drop quickly as the shorter-lived isotopes
decay.  

The procedure that must be followed, should personnel require access to the Controlled Areas,
is that a Radiation Safety Officer must conduct a survey of the areas where access is
desired, and establish that radiation levels are sufficiently low to allow this access.  
This officer will also cordon off areas where activity levels are too high to access, 
and can also establish limits on the amount of time a person can work at any location inside 
the Controlled Area.

Access must be carefully monitored, and all personnel working in such areas must have
received proper training, must have radiation-monitoring devices (badges), and must be
properly logged in and out of the controlled areas.  Access will be controlled with lock-out, tag out procedures interlocked to to the beam producing sub-systems of the cyclotron.

\subsection{Controlled Areas}

There are several types of Controlled Areas, mainly distinguished by the 
level of radiation.  
When beam is not running, the highest levels will likely be found
 inside the cyclotron shielding vault, and
the target vault.  The radiation in these areas will be primarily gammas, from
the cyclotron extraction region and the filters of the primary coolant circuit
in the target.  These will require the highest level of access control.

Other areas, such as the electrical substation and power supply areas around the
cyclotron and the beam transport line, are likely to experience much lower radiation
levels during Beam-Off times, but under any circumstances monitoring of radiation
levels is important to ensure personnel access is safe.

\subsubsection{Siting: Original location}

As discussed in section 2.1, two different siting options for IsoDAR have been considered. 
The first one utilizes existing passageways and caverns, while the second, preferred option, is 
in a separate location away from the areas accessed by KamLAND personnel.

For the first scenario, during Beam-On times the passageway leading to the 
KamLAND dome must be off limits,
because of its proximity to the transport line.  The wall separating this passageway from
the beam transport line may provide some shielding, to allow safe access during Beam-Off
times, but is definitely inadequate to provide protection while the beam is on.

The primary gate into the KamLAND area, next to the cooling water intake from the
underground stream, should probably be the boundary of the exclusion area.
During Beam-On time this gate should be closed, and interlocked, so no personnel can
enter the area.

When beam is off, and after suitable survey by the Radiation Protection Officer, access
can be granted to the principal KamLAND areas, and only the cyclotron shielding vault and 
target vaults remaining unaccessible pending more complete surveys.

\subsubsection{Siting:  Preferred location}

Access to the IsoDAR equipment, under the preferred scenario, is via an extension
of the drift currently housing the first liquid scintillator purification system, as seen
in Figure~\ref{new-site}.  

The boundary of the Controlled Area would be a gate located along this drift.  
Access control procedures would remain the same as above, 
but would affect only the areas beyond this gate, and so 
would have virtually no effect on KamLAND personnel and their normal activities.

It is even most likely that personnel could access the KamLAND dome
while beam is on.  

\section{Monitoring Instrumentation and Interlock Systems}

The specifications for radiation-safety instrumentation are quite well
established, and systems in existence in laboratories around the world,
and particularly in Japan, are excellent models for the systems that
need to be implemented at IsoDAR.



\clearpage

\chapter{Conclusions}

This document shows that the requirements and logistics for 
transporting, installing, and operating the IsoDAR experiment
at the KamLAND site are feasible, but must be carefully planned.
This CDR provides an assessment of the space needed for the
components, shielding, and support electrical
equipment, which will be refined for the PDR.   At either proposed
site, cavern enlargement will be required.  Techniques for gentle rock
removal are already used by the mining company.

Two possible siting options have been developed:  one integrated into the existing
drifts and caverns; a second involving excavation of a new
cavern on the opposite side of the KamLAND detector.  This second site offers
advantages, and so is the preferred site.  However, a further study is
needed to fully demonstrate that the additional costs for removing a larger volume
of rock, are offset by the operational advantages of having an independent site.

Shielding specifications have been set from the very stringent
requirements to keep activation of rock to extremely low levels.
We have found that shielding to meet these requirements is
indeed possible, and have established upper-bounds on the thickness
and materials to use in these shielding configurations.

Transport of equipment will require breaking all systems down into
components of  about 10 tons in weight.   This limit is set by the condition
of the roadbed on the secondary roads leading to the mine, and the
truck size that can navigate these narrow, windy roads.  
Plans for breaking the cyclotron into small parts have already been included in the
Technical Facility CDR.

A warehouse will be needed about halfway between Toyama and the mine,
with sufficient area to stage the number of pallets corresponding to a shipping
plan that must be developed.
Staging areas at the KamLAND site must also be identified.   Within
the proposed caverns, the plans include
adequate space for lay-down and assembly of the components.
Rigging can be accomplished by a number of means, 
including heavy-duty overhead hoists, with rails that are anchored
into the rock of the ceiling of the caverns.   These techniques are
already used in the mine.

Utilities needs have been established, and plans are being developed for adding the required
electrical, cooling and ventilation systems.

Radiation protection considerations have been carefully studied, with assistance from
the very experienced and able staff of RIKEN.  Shielding calculations for the target area
have reached a high level of maturity for mitigating rock activity, and are beginning for
evaluation of acceptable background levels in the KamLAND detector.
Shielding calculations are currently ongoing for the cyclotron and 
transport areas.  These calculations, performed with GEANT4 by the IsoDAR team,
are being reproduced and verified by RIKEN experts using the PHITS code.

Plans have been developed for handling of radioactive material for maintenance
activities, and for isolation
and monitoring for possible air and ground-water contamination.

The IsoDAR team acknowledges the extremely 
valuable contributions and support of the Radiation Protection office of RIKEN, Yoshitomo
Uwamino and Kanenobu Tanaka.

In summary, it appears that there are reasonably good solutions to all the issues
that have been identified.

This is a ``Conceptual Design" in that we present possible plans that can meet 
all the site requirements. These must be carefully developed by an engineering
team, to produce a more mature set of project plans for deployment 
and operation of the IsoDAR experiment, as part of the future PDR for IsoDAR$@$KamLAND.

\end{document}